\documentclass[aps,prd,onecolumn,groupedaddress,showpacs,nofootinbib,amssymb]{revtex4}
\usepackage[dvips]{graphicx}
\usepackage{amssymb}
\usepackage{amsmath}
\usepackage{graphicx,color}
\usepackage{amsfonts}
\usepackage{bm}
\usepackage{cancel}
\usepackage{comment}

\newcommand\e{\mathrm{e}}

\allowdisplaybreaks[4] \tolerance=5000

\begin{document}

\title{Viable Inflationary Models in a Ghost-free Gauss-Bonnet Theory of Gravity}
\author{Shin'ichi~Nojiri,$^{1,2}$\,\thanks{nojiri@gravity.phys.nagoya-u.ac.jp}
S.~D.~Odintsov,$^{3,4}$\,\thanks{odintsov@ieec.uab.es}
V.~K.~Oikonomou,$^{5,6,7}$\,\thanks{v.k.oikonomou1979@gmail.com}
N.~Chatzarakis,$^{5}$ Tanmoy~Paul$^8$\thanks{pul.tnmy9@gmail.com}}
\affiliation{ $^{1)}$ Department of Physics, Nagoya University,
Nagoya 464-8602, Japan \\
$^{2)}$ Kobayashi-Maskawa Institute for the Origin of Particles
and the Universe, Nagoya University, Nagoya 464-8602, Japan \\
$^{3)}$ ICREA, Passeig Luis Companys, 23, 08010 Barcelona, Spain\\
$^{4)}$ Institute of Space Sciences (IEEC-CSIC) C. Can Magrans
s/n,
08193 Barcelona, Spain\\
$^{5)}$ Department of Physics, Aristotle University of
Thessaloniki, Thessaloniki 54124,
Greece\\
$^{6)}$ International Laboratory for Theoretical Cosmology, Tomsk
State University of Control Systems
and Radioelectronics (TUSUR), 634050 Tomsk, Russia\\
$^{7)}$ Tomsk State Pedagogical University, 634061 Tomsk, Russia\\
$^{8)}$ Department of Theoretical Physics,\\
Indian Association for the Cultivation of Science,\\
2A $\&$ 2B Raja S.C. Mullick Road,\\
Kolkata - 700 032, India }

\tolerance=5000

\begin{abstract}
In this work we investigate the inflationary phenomenological
implications of a recently developed ghost-free Gauss-Bonnet
theory of gravity. The resulting theory can be viewed as a scalar
Einstein-Gauss-Bonnet theory of gravity, so by employing the
formalism for cosmological perturbations for the latter theory, we
calculate the slow-roll indices and the observational indices, and
we compare these with the latest observational data. Due to the
presence of a freely chosen function in the model, in principle
any cosmological evolution can be realized, so we specify the
Hubble rate and the freely chosen function and we examine the
phenomenology of the model. Specifically we focus on de Sitter,
quasi-de Sitter and a cosmological evolution in which the Hubble
rate evolves exponentially, with the last two being more realistic
choices for describing inflation. As we demonstrate, the
ghost-free model can produce inflationary phenomenology compatible
with the observational data. We also briefly address the stability
of first order scalar and tensor cosmological perturbations, for
the exponential Hubble rate, and as we demonstrate, stability is
achieved for the same range of values of the free parameters that
guarantee the phenomenological viability of the models.
\end{abstract}

\pacs{04.50.Kd, 95.36.+x, 98.80.-k, 98.80.Cq,11.25.-w}

\maketitle

\section{Introduction \label{SecI}}

Nearly forty years ago, three of the major problems in
contemporary cosmology, namely the Horizon Problem, the Flatness
Problem and the Magnetic-Monopoles Problem, have been given a
successful solution in the context of the inflationary scenario.
This scenario was firstly proposed in Ref.~\cite{Guth:1980zm} and
was further developed in Ref.~\cite{Linde:1981mu,Albrecht:1982wi}.
According to the inflationary scenario, merely fractions of
seconds after the Big Bang, the spatial coordinates of the
Universe expanded exponentially. An expansion of this sort is
supposed to last from about $10^{-36}$sec to $10^{-15}$sec and the
size of the Universe is increased by a factor of $10^{26}$. The
nature of this scenario is rather bizarre for classical cosmology,
since traditional Big Bang Friedmann-Robertson-Walker (FRW) models
do not match the fast evolution of the Universe
\cite{Linde:2007fr,Gorbunov:2011zzc,Lyth:1998xn}. The first
approximation is to consider the expansion as the de Sitter phase
of the Universe. The standard approach to achieve the de Sitter
inflationary phase in cosmology is to use scalar fields, and many
of the initial models of inflation made use of the scalar field
formalism.

However, it is also possible to produce an inflationary phase of
the Universe in the context of modified gravity, see
Refs.~\cite{Nojiri:2017ncd,Nojiri:2010wj,Nojiri:2006ri,Capozziello:2011et,
Capozziello:2010zz,delaCruzDombriz:2012xy,Olmo:2011uz} for reviews
on this. In fact, the first model of $f(R)$ which remains viable
up to date is the Starobinsky model \cite{Starobinsky:1982ee}, and
ever since many models have been developed in various forms of
modified gravity
\cite{Nojiri:2017ncd,Nojiri:2010wj,Nojiri:2006ri,Capozziello:2011et,
Capozziello:2010zz,delaCruzDombriz:2012xy,Olmo:2011uz}. In all the
modified gravities the key element is that geometric terms are
included in the gravitational Lagrangian, which are absent in the
Einstein-Hilbert gravity. These terms may dominate the Universe's
evolution at early times or even at late times. Such models may
include additional curvature terms, namely the $f(R)$ theories,
torsional terms namely the teleparallel $f(T)$ theories, or the
Gauss-Bonnet modified gravities $f(\mathcal{G})$ theories, as well
as the generalized $f(R,\mathcal{G})$ theories (see
\cite{Nojiri:2017ncd,Nojiri:2010wj,Nojiri:2006ri,Capozziello:2011et,
Capozziello:2010zz,delaCruzDombriz:2012xy,Olmo:2011uz}). Such
theoretical formulations of gravity are able to model both the
early-time expansion and the late-time acceleration, see for
example \cite{Nojiri:2003ft}.

Recently it was demonstrated how ghosts may disappear from the
Gauss-Bonnet modified gravity theories in general background
\cite{Nojiri:2018ouv}. In \cite{Nojiri:2018ouv}, we have
considered the perturbation from the general background with
matter and we have chosen the coordinate system where $g_{tt}=-1$
and $g_{ti}=g_{it}= 0$ $\left( i=1,2,3 \right)$. Then after
eliminating the perturbation of the scalar field, we have shown
that the perturbed equations do not include higher than second
order derivatives of the metric and therefore there no ghosts in
the general background. Actually in Ref.~\cite{Nojiri:2018ouv} it
was thoroughly investigated how ghost degrees of freedom may occur
in $f(\mathcal{G})$ and $f(R,\mathcal{G})$ theories, and how the
theory should be modified in order for these ghost degrees of
freedom to disappear at the equations of motion level. In this
work we shall be interested on the inflationary aspects of ghost
free $f(\mathcal{G})$ gravity theory developed in
\cite{Nojiri:2018ouv}. The ghost free $f(\mathcal{G})$ gravity
contains a scalar field and the resulting theory can be treated as
an effective scalar Einstein-Gauss-Bonnet theory. By employing the
slow-roll approximation, we shall calculate the slow-roll indices
of the resulting theory and the corresponding observational
indices of inflation, and we shall confront the results with the
latest Planck \cite{Akrami:2018odb} and BICEP2/Keck-Array data
\cite{Array:2015xqh}. Due to the freedom offered by the theory by
construction, we shall demonstrate that the resulting theory can
be viable. This is due to the presence of the Lagrange multiplier
terms, as we show. Thus by treating the theory as the effective
Einstein-Gauss-Bonnet theory, we shall fix initially the Hubble
rate and the function $h(\chi)$, which is the coupling of the
Gauss-Bonnet term, and we shall investigate which theory can
realize the given evolution. Accordingly, we shall investigate the
phenomenological viability of the models, by calculating the
observational indices and by directly confronting the theory with
the latest Planck \cite{Akrami:2018odb} data. We shall use three
different types of cosmological evolutions, namely the de Sitter,
the quasi-de Sitter and an exponential type of cosmological
evolution. As we shall demonstrate, in the last two cases, the
viability with the observational data can be achieved by
appropriately restricting the values of the free parameters. In
addition, we shall use another reconstruction approach, in the
context of which we shall fix the Hubble rate and the scalar
potential, instead of the function $h(\chi)$, and we shall perform
the same analysis in order to test the phenomenological viability
of the model. As we demonstrate, the viability of the theory is
also achieved in this case too, by appropriately constraining the
values of the free parameters. Finally, we examine the stability
of first order scalar and tensor perturbations, for the
exponential cosmological evolution, and as we demonstrate these
perturbations are stable for the same range of values of the free
parameters, for which the phenomenological viability of the model
is achieved.

This paper is organized as follows: In section \ref{SecII} we
briefly review the essential features of the ghost free
$f(\mathcal{G})$ gravity, in section \ref{SecIII} we present the
inflationary dynamics formalism of $f(\mathcal{G})$ gravity which
we shall use in the rest of the paper. Accordingly in section
\ref{SecIV} we investigate how this formalism can be applied for
the case that the de Sitter evolution is chosen, by also choosing
the functional form of the function coupling function $h(\chi)$.
Accordingly, in section \ref{SecV} we discuss the case of a
quasi-de Sitter evolution. In section \ref{SecVI}, an exponential
cosmological evolution is studied in detail in the same context as
in the previous sections, and also the stability of the first
order perturbations is investigated too. Finally, the conclusions
follow in the end of the paper.

\section{Essential Features of Ghost-free $f(\mathcal{G})$ Gravity \label{SecII}}

In this section we shall recall the essential features of the
ghost free $f(\mathcal{G})$ gravity developed in
Ref.~\cite{Nojiri:2018ouv}. The whole ghost-free construction
scheme is based on introducing a Lagrange multiplier $\lambda$ in
the standard $f(\mathcal{G})$ gravity action, so the ghost-free
action is the following,
\begin{equation}
\label{FRGBg19} S=\int d^4x\sqrt{-g} \left(\frac{1}{2\kappa^2}R +
\lambda \left( \frac{1}{2} \partial_\mu \chi \partial^\mu \chi +
\frac{\mu^4}{2} \right)
 - \frac{1}{2} \partial_\mu \chi \partial^\mu \chi
+ h\left( \chi \right) \mathcal{G} - V\left( \chi \right) +
\mathcal{L}_\mathrm{matter}\right)\, ,
\end{equation}
where $\mu$ is a mass-dimension one constant. Upon variation with
respect to the Lagrange multiplier $\lambda$, we obtain the
following constraint equation,
\begin{equation}
\label{FRGBg20} 0=\frac{1}{2} \partial_\mu \chi \partial^\mu \chi
+ \frac{\mu^4}{2} \, .
\end{equation}
Effectively, the kinetic term is a constant, so it can be absorbed
in the scalar potential in the following way,
\begin{equation}
\label{FRGBg21} \tilde V \left(\chi\right) \equiv \frac{1}{2}
\partial_\mu \chi \partial^\mu \chi + V \left( \chi \right)
= - \frac{\mu^4}{2} + V \left( \chi \right) \, ,
\end{equation}
and in effect, the action of Eq.~(\ref{FRGBg19}) is rewritten as,
\begin{equation}
\label{FRGBg22} S=\int d^4x\sqrt{-g} \left(\frac{1}{2\kappa^2}R +
\lambda \left( \frac{1}{2} \partial_\mu \chi \partial^\mu \chi +
\frac{\mu^4}{2} \right) + h\left( \chi \right) \mathcal{G}
 - \tilde V\left( \chi \right) + \mathcal{L}_\mathrm{matter}\right)
\, .
\end{equation}
The equations of motion for the action (\ref{FRGBg22}), are
(\ref{FRGBg20}) and the following,
\begin{align}
\label{FRGBg23} 0 =& - \frac{1}{\sqrt{-g}} \partial_\mu \left(
\lambda g^{\mu\nu}\sqrt{-g}
\partial_\nu \chi \right)
+ h'\left( \chi \right) \mathcal{G} - {\tilde V}'\left( \chi \right) \, , \\
\label{FRGBg24} 0 =& \frac{1}{2\kappa^2}\left(- R_{\mu\nu} +
\frac{1}{2}g_{\mu\nu} R\right) + \frac{1}{2} T_{\mathrm{matter}\,
\mu\nu}
 - \frac{1}{2} \lambda \partial_\mu \chi \partial_\nu \chi
 - \frac{1}{2}g_{\mu\nu} \tilde V \left( \chi \right)
+ D_{\mu\nu}^{\ \ \tau\eta} \nabla_\tau \nabla_\eta h \left( \chi
\right)\, ,
\end{align}
Upon multiplication of Eq.~(\ref{FRGBg24}) with $g^{\mu\nu}$, we
get,
\begin{equation}
\label{FRGBg24A} 0 = \frac{R}{2\kappa^2} + \frac{1}{2}
T_\mathrm{matter} + \frac{\mu^4}{2} \lambda - 2 \tilde V \left(
\chi \right) - 4 \left( - R^{\tau\eta} + \frac{1}{2} g^{\tau\eta}
R \right) \nabla_\tau \nabla_\eta h \left( \chi \right) \, ,
\end{equation}
By solving  Eq.~(\ref{FRGBg24A}) with respect to $\lambda$, we
get,
\begin{equation}
\label{FRGBg24AB} \lambda = - \frac{2}{\mu^4} \left(
\frac{R}{2\kappa^2} + \frac{1}{2} T_\mathrm{matter}
 - 2 \tilde V \left( \chi \right) - 4 \left( - R^{\tau\eta}
+ \frac{1}{2} g^{\tau\eta} R \right) \nabla_\tau \nabla_\eta h
\left( \chi \right) \right) \, .
\end{equation}
Let us now see how the equations of motion become if the metric
background is a flat Friedmann-Robertson-Walker (FRW), with line
element,
\begin{equation}
\label{FRWmetric} ds^2 = - dt^2 + a(t)^2 \sum_{i=1,2,3} \left(
dx^i \right)^2 \, .
\end{equation}
Assuming that the functions $\lambda$ and $\chi$ are only cosmic
time dependent, and also that no matter fluids are present, that
is, $T_{\mathrm{matter}\, \mu\nu} =0$, Eq.~(\ref{FRGBg20}) has the
following simple solution,
\begin{equation}
\label{frgdS4} \chi = \mu^2 t \, .
\end{equation}
Hence, the $(t,t)$ and $(i,j)$ components of Eq.~(\ref{FRGBg24})
can be written,
\begin{align}
\label{FRGFRW1} 0 = & - \frac{3H^2}{2\kappa^2}
 - \frac{\mu^4 \lambda}{2} + \frac{1}{2} \tilde V \left( \mu^2 t \right)
 - 12 \mu^2 H^3 h' \left( \mu^2 t \right) \, , \\
\label{FRGFRW2} 0 = & \frac{1}{2\kappa^2} \left( 2 \dot H + 3 H^2
\right)
 - \frac{1}{2} \tilde V \left( \mu^2 t \right)
+ 4 \mu^4 H^2 h'' \left( \mu^2 t \right) + 8 \mu^2 \left( \dot H +
H^2 \right) H h' \left( \mu^2 t \right) \, ,
\end{align}
and in addition from Eq.~(\ref{FRGBg23}) we get,
\begin{equation}
\label{FRGFRW3} 0 = \mu^2 \dot\lambda + 3 \mu^2 H \lambda + 24 H^2
\left( \dot H + H^2 \right) h'\left( \mu^2 t \right)
 - {\tilde V}'\left( \mu^2 t \right) \, .
\end{equation}
By solving Eq.~(\ref{FRGFRW1}) with respect to $\lambda$ we get,
\begin{equation}
\label{FRGFRW4} \lambda = - \frac{3 H^2}{\mu^4 \kappa^2} +
\frac{1}{\mu^4} \tilde V \left( \mu^2 t \right) - \frac{24}{\mu^2}
H^3 h' \left( \mu^2 t \right) \, .
\end{equation}
It is easy to see that by combining Eqs.~(\ref{FRGFRW4}) and
(\ref{FRGFRW3}), we easily obtain Eq.~(\ref{FRGFRW2}). Also by
solving Eq.~(\ref{FRGFRW2}) with respect to the scalar potential
$\tilde V \left( \mu^2 t \right)$, we get,
\begin{equation}
\label{FRGFRW7} \tilde V \left( \mu^2 t \right) =
\frac{1}{\kappa^2} \left( 2 \dot H + 3 H^2 \right) + 8 \mu^4 H^2
h'' \left( \mu^2 t \right) + 16 \mu^2 \left( \dot H + H^2 \right)
H h' \left( \mu^2 t \right) \, .
\end{equation}
Hence, for an arbitrarily chosen function $h(\chi (t))$, and with
the potential $\tilde V \left( \chi \right)$ being equal to,
\begin{equation}
\label{FRGFRW8} \tilde V \left( \chi \right) = \left[
\frac{1}{\kappa^2} \left( 2 \dot H + 3 H^2 \right) + 8 \mu^4 H^2
h'' \left( \mu^2 t \right) + 16 \mu^2 \left( \dot H + H^2 \right)
H h' \left( \mu^2 t \right) \right]_{t=\frac{\chi}{\mu^2}}\, ,
\end{equation}
then we can realize an arbitrary cosmology corresponding to a
given Hubble rate $H(t)$. Finally, the functional form of the
Lagrange multiplier is equal to,
\begin{equation}
\label{FRGFRW4B} \lambda = \frac{2 \dot H}{\mu^4 \kappa^2} + 8 H^2
h'' \left( \mu^2 t \right) + \frac{8}{\mu^2} \left( 2 \dot H - H^2
\right) H h' \left( \mu^2 t \right) \, .
\end{equation}
The resulting theory with Lagrangian (\ref{FRGBg22}) is a form of
the scalar Einstein-Gauss-Bonnet gravity and in the next section
we shall extensively discuss the inflationary dynamics of this
model. The presence of the arbitrary function $h(\chi)$ provides
us with the freedom of realizing several viable cosmologies.

\section{Inflationary Dynamics of the Ghost-free
$f(\mathcal{G})$ Model \label{SecIII}}

As we already mentioned, the ghost-free $f(\mathcal{G})$ model of
Eq.~(\ref{FRGBg22}) is a sort of scalar Einstein-Gauss-Bonnet
model
\cite{Nojiri:2006je,Cognola:2006sp,Nojiri:2005vv,Nojiri:2005jg,Nojiri:2007te,
Bamba:2014zoa,Yi:2018gse,Guo:2009uk,Guo:2010jr,Jiang:2013gza,Koh:2014bka,
Koh:2016abf,Kanti:2015pda,vandeBruck:2017voa,Kanti:1998jd,Nozari:2017rta,
Chakraborty:2018scm,Odintsov:2018zhw}, the cosmological
perturbations of which were studied in Ref.~\cite{Hwang:2005hb}.
In this section we shall use the formalism, notation and results
of Ref.~\cite{Hwang:2005hb}, and we shall calculate the spectral
index of primordial curvature perturbations and the
tensor-to-scalar ratio for the model (\ref{FRGBg22}), by
specifying the functional form of $h(\chi)$ and the Hubble rate.
Then, by replacing the cosmic time with the $e$-foldings number,
we shall express all the observational and slow-roll indices as
functions of the $e$-foldings number, and we shall put the
phenomenology of the model into test by confronting the resulting
theory with the latest observational data.

We begin by defining the functions $Q_{i}(\chi)$ (see
\cite{Hwang:2005hb} for more details), as follows,
\begin{align}
\label{qfunctionstime} & Q_{a}(\chi) =  8 \dot{h}(\chi) H^{2}\, ,
\quad Q_{b}(\chi) = 16 \dot{h}(\chi) H \, ,\quad
Q_{c}(\chi) = Q_{d}(\chi) = 0 \, ,\nonumber \\
& Q_{e}(\chi) = 32 \dot{h}(\chi) \dot{H} \, ,\quad Q_{f}(\chi) =
-16 \left( \ddot{h}(\chi) - \dot{h}(\chi) H \right)\, , \quad
Q_{t}(\chi) = 1 + 8\dot{h}(\chi) H\, ,
\end{align}
where $H$ is the Hubble rate, $H\equiv \dot a/a$. In addition, the
wave speeds $c_{A}$ and $c_{T}$ become,
\begin{equation*}
c_{A}^{2} = \dfrac{X \frac{\partial f}{\partial X} + \frac{3
\dot{F}^{2}}{2 F}}{X \frac{\partial f}{\partial X} + 2 X^{2}
\frac{\partial^{2}f}{\partial X^{2}} + \frac{3 \dot{F}^{2}}{2
F}}\, , \quad c_{T}^{2} = 1 - \dfrac{Q_{f}}{2 F + Q_{b}} \, ,
\end{equation*}
where $X = -\dfrac{1}{2} \dot{\chi}^{2}$ , $\dfrac{\partial
f}{\partial X} = \dfrac{\lambda}{2}$, $\dfrac{\partial^{2}
f}{\partial X^{2}} = 0$ and $F=1$ in our case. Note that $c_A$ is
the wave speed of the perturbed field in the context of the
perturbed FRW metric, and $c_T$ is the sound speed. For more
details on this we refer the reader to \cite{Hwang:2005hb}. The
definition of the wave speeds is for the general Gauss-Bonnet
corrected $f(R,\chi)$ theory with $F=\frac{\partial f}{\partial
R}$, but in our case $f(R,\chi)=R$ and $F=1$. Also the waves
speeds are affected from the Gauss-Bonnet coupling via the
functions $Q_f$ and $Q_b$ which in our case have the form
(\ref{qfunctionstime}). As a result, the two wave speeds are
further simplified with the wave speed of the perturbed field
$c_A$ being trivial as in the classical case,
\begin{equation}
\label{wavespeed1} c_{A}^{2} = 1 \, ,
\end{equation}
while the wave speed of the gravitational waves in non-trivial,
\begin{equation}
\label{wavespeedtime} c_{T}^{2} = 1 + \dfrac{16 \left(
\ddot{h}(\chi) - \dot{h}(\chi) H \right) }{2 + 16 \dot{h}(\chi)
H}\, .
\end{equation}
In order to calculate the slow-roll parameters, we first need to
determine the function $E(R,\chi,X)$ which is defined as follows
\cite{Hwang:2005hb},
\begin{equation}
\label{Etime} E(R,\chi,X) = \dfrac{F(R,\chi)}{\dot{\chi}} \left(
\omega(\chi) \dot{\chi}^{2} + 3 \dfrac{ \left( \dot{F}(R,\chi) +
Q_{a} \right)^{2} }{2 F(R,\chi) + Q_{b}} \right) = - \lambda
\dot{\chi} + \dfrac{192 \dot{h}(\chi)^{2} H^{4}}{2 \dot{\chi} + 16
\dot{\chi} \dot{h}(\chi) H} \, .
\end{equation}
The slow-roll parameters are defined as follows
\cite{Hwang:2005hb}
\begin{align}
\label{slowrolltime} & \epsilon_{1} = \dfrac{\dot{H}}{H^{2}}\, ,
\quad \epsilon_{2} = \dfrac{\ddot{\chi}}{H \dot{\chi}} = 0\, ,
\quad \epsilon_{3} = \dfrac{1}{2} \dfrac{\dot{F}(R,\chi)}{H
F(R,\chi)} = 0 \, , \quad
\epsilon_{4} = \dfrac{1}{2} \dfrac{\dot{E}(R,\chi,X)}{H E(R,\chi,X)} \, , \nonumber \\
& \epsilon_{5} = \dfrac{\dot{F} + Q_{a}}{H \left( 2 F(R,\chi) +
Q_{b} \right) } = \dfrac{4 \dot{h}(\chi) H^{2}}{H \left( 1 + 8
\dot{h}(\chi) H \right) } \, , \quad \epsilon_{6} =
\dfrac{\dot{Q}_{t}}{2 H Q_{t}} = \dfrac{4 \ddot{h}(\chi) H + 4
\dot{h}(\chi) \dot{H}}{ H \left( 1 + 8 \dot{h}(\chi) H \right) }
\, .
\end{align}
The two spectral indices, for scalar and for tensor perturbations
in the inflationary era respectively, are defined using the
slow-roll parameters \cite{Hwang:2005hb},
\begin{equation}
\label{spectralind} n_{S} = 1 + 2 \dfrac{\epsilon_{1} -
\epsilon_{2} + \epsilon_{3}
 - \epsilon_{4}}{1 + \epsilon_{1}} \, , \quad
n_{T} = 2 \dfrac{\epsilon_{1} - \epsilon_{6}}{1 + \epsilon_{1}} \,
.
\end{equation}
Finally, the tensor-to-scalar ratio  is equal to
\cite{Hwang:2005hb},
\begin{equation}
\label{tensorscalar} r = 4 \left| \left[ \epsilon_{1} -
\epsilon_{3} - \dfrac{1}{4 F(R,\chi) } \left( \dfrac{1}{H^{2}} (2
Q_{c} + Q_{d}) - \dfrac{1}{H} Q_{e} + Q_{f} \right) \right]
\dfrac{1}{1 + \frac{Q_{b}}{2 F(R,\chi)}} \left(
\dfrac{c_{A}}{c_{T}} \right)^{3} \right| \, .
\end{equation}
The above expressions of the parameters for the slow-roll
inflationary dynamics, are in fact functions of the cosmic time,
$t$. However, such a description is not sufficient for our study,
since the preferable variable to perfectly quantify the evolution
during the inflationary era is the $e$-foldings number, $N$. So we
need to transform the above relations with respect to the
$e$-foldings numbers. At first, we consider a given Hubble
expansion rate for the inflationary era, as a function of time, $H
= H(t)$. The $e$-foldings number is defined as
\begin{equation}
\label{efold} N = \int_{t_{i}}^{t_{f}} H(t) dt \, ,
\end{equation}
where $t_{i}$ is the initial and $t_{f}$ the final moments of
inflation. Considering a given initial moment for inflation,
$t_{i} \in [ 0, 10^{-36} ]$, and an unspecified final moment, $t$,
the $e$-foldings number is obtained via Eq.~\ref{efold} as a
function of time, $N = N(t)$. Supposing this function is
reversible, time is also given as a function of the $e$-foldings
number, $t = t(N)$. Consequently, the first- and the second-order
derivatives with respect to time, are transformed into first- and
second-order derivatives with respect to the $e$-foldings number,
as follows,
\begin{equation}
\label{derivatives} \dfrac{d}{dt} = \dfrac{dN}{dt} \dfrac{d}{dN} =
H(N) \dfrac{d}{dN}\, , \quad \dfrac{d^{2}}{dt^{2}} = \left(
\dfrac{dN}{dt} \right)^{2} \dfrac{d^{2}}{dN^{2}} + \dfrac{dN}{dt}
\dfrac{dH}{dN} \, , \quad  \dfrac{d}{dN} = H(N)^{2}
\dfrac{d^{2}}{dN^{2}} + H(N) \dfrac{dH}{dN} \dfrac{d}{dN} \, .
\end{equation}
Since the scalar field, $\chi = \chi(t)$ is a function of time,
its potential, $\tilde{V}(\chi) = \tilde{V}(\chi(t))$, and the
Lagrange multiplier, $\lambda = \lambda(t)$, the coupling
function, $h(\chi) = h(\chi(t))$, as well as the Ricci scalar, the
Gauss-Bonnet invariant and the function $E(R,\chi) = E \left(
R(t),\chi(t) \right)$ are also functions of the cosmic time. As a
result, they can all be rewritten with respect to the $e$-foldings
number. Furthermore, the functions $Q_{i}(\chi)$ are also
transformed, taking the following forms,
\begin{align}
\label{qfunctionsefold} Q_{a} (N) &= 8 H(N)^{2} h'(N) \, , \quad
Q_{b} (N) = 16 H(N)^{2} h'(N) \, , \quad Q_{c} (N) = Q_{d} (N) = 0
\, , \quad
Q_{e} (N) = 32 H(N)^{2} H'(N) h'(N) \, , \nonumber \\
Q_{f} (N) &= -16 \left( H(N)^{2} h''(N) + H(N) H'(N) h'(N) -
H(N)^{2} h'(N) \right) \, , \quad Q_{t} (N) = 1 + 8 H(N)^{2} h'(N)
\, ,
\end{align}
where the prime denotes differentiation with respect to the
$e$-foldings number. In the same manner, we may redefine the wave
speed for the gravitational waves,
\begin{equation}
\label{wavespeedefold} c_{T}^{2} = 1 - \dfrac{Q_{f}(N)}{2 + Q_{b}
(N)} = 1 + \dfrac{8 \left( H(N)^{2} h''(N) + H(N) H'(N) h'(N) -
H(N)^{2} h'(N) \right)} {1 + 8 H(N)^{2} h'(N)} \, .
\end{equation}
The next step is to express the slow-roll parameters,
$\epsilon_{i}$, with respect to the $e$-foldings number, and the
resulting expressions are,
\begin{align}
\label{slowrollefold} \epsilon_{1} (N) &= \dfrac{H'(N)}{H(N)} \, ,
\quad \epsilon_{2} (N) = \dfrac{\chi''(N)}{\chi'(N)} +
\dfrac{H'(N)}{H(N)} = 0 \, , \quad \epsilon_{3} (N) = \dfrac{1}{2}
\dfrac{F'(N)}{F(N)} = 0 \, , \quad
\epsilon_{4} (N) = \dfrac{1}{2} \dfrac{E'(N)}{E(N)} \, , \nonumber \\
\epsilon_{5} (N) &= \dfrac{Q_{a}(N)}{H(N) \left( 2 + Q_{b}(N)
\right)}
= \dfrac{ 4 H(N) h'(N) }{ 1 + 8 H(N)^{2} h'(N) } \, , \nonumber \\
\epsilon_{6} (N) &= \dfrac{Q_{t}'(N)}{Q_{t}(N)} = \dfrac{ H(N)
\left( 16 H'(N) h'(N) + 8 H(N) h''(N) \right) } { 1 + 8 H(N)^{2}
h'(N) } \, .
\end{align}
Through these, the spectral indices and the tensor-to-scalar ratio
are directly calculated with respect to the $e$-foldings number,
using Eqs.~(\ref{spectralind}) and (\ref{tensorscalar}).

What remains  is to define a specific coupling function,
$h(\chi)$, as well as the Hubble rate for the cosmological FRW
background, and also to calculate the spectral indices and the
tensor-to-scalar ratio and compare our results with that of the
latest Planck \cite{Akrami:2018odb} and BICEP2/Keck-Array
\cite{Array:2015xqh} observations. With regard to the coupling
function, we shall assume that it has either exponential or
power-law forms, while with regard to the Hubble rate, we shall
firstly assume the de Sitter evolution for a warm up study, and
finally we shall assume the quasi-de Sitter evolution.

\section{The Case of de Sitter Background Evolution \label{SecIV}}

In the de Sitter case, the Hubble rate is constant as a function
of the cosmic time,
\begin{equation}
\label{desitter-hubble} H(t) = H_{0} \, ,
\end{equation}
therefore the $e$-foldings number and the cosmic time are related
as follows,
\begin{equation}
\label{desitter-time} t = \dfrac{N}{H_{0}} \ .
\end{equation}
As a result, the Ricci scalar and the Gauss-Bonnet invariant are
both constant,
\begin{equation}
\label{desitter-curvature} R = 12 H_{0}^{2}\, , \quad \mathcal{G}
= 24 H_{0}^{4} \, .
\end{equation}
Finally, the scalar field given by Eq.~(\ref{frgdS4}), takes the
following form,
\begin{equation}
\label{desitter-scalar} \chi(N) = \dfrac{\mu^{2}}{H_{0}} N \, .
\end{equation}
Using, Eqs.~(\ref{desitter-hubble}), (\ref{desitter-time}) and
(\ref{desitter-scalar}) and in addition a specific form for the
function $h(\chi)$, we can calculate the slow-roll indices and the
observational indices for the de Sitter evolution cosmology.

\subsection{A power-law coupling function, $h(\chi) = \gamma \chi^{b}$}

\begin{figure}[h!]
\centering
\includegraphics[width=18pc]{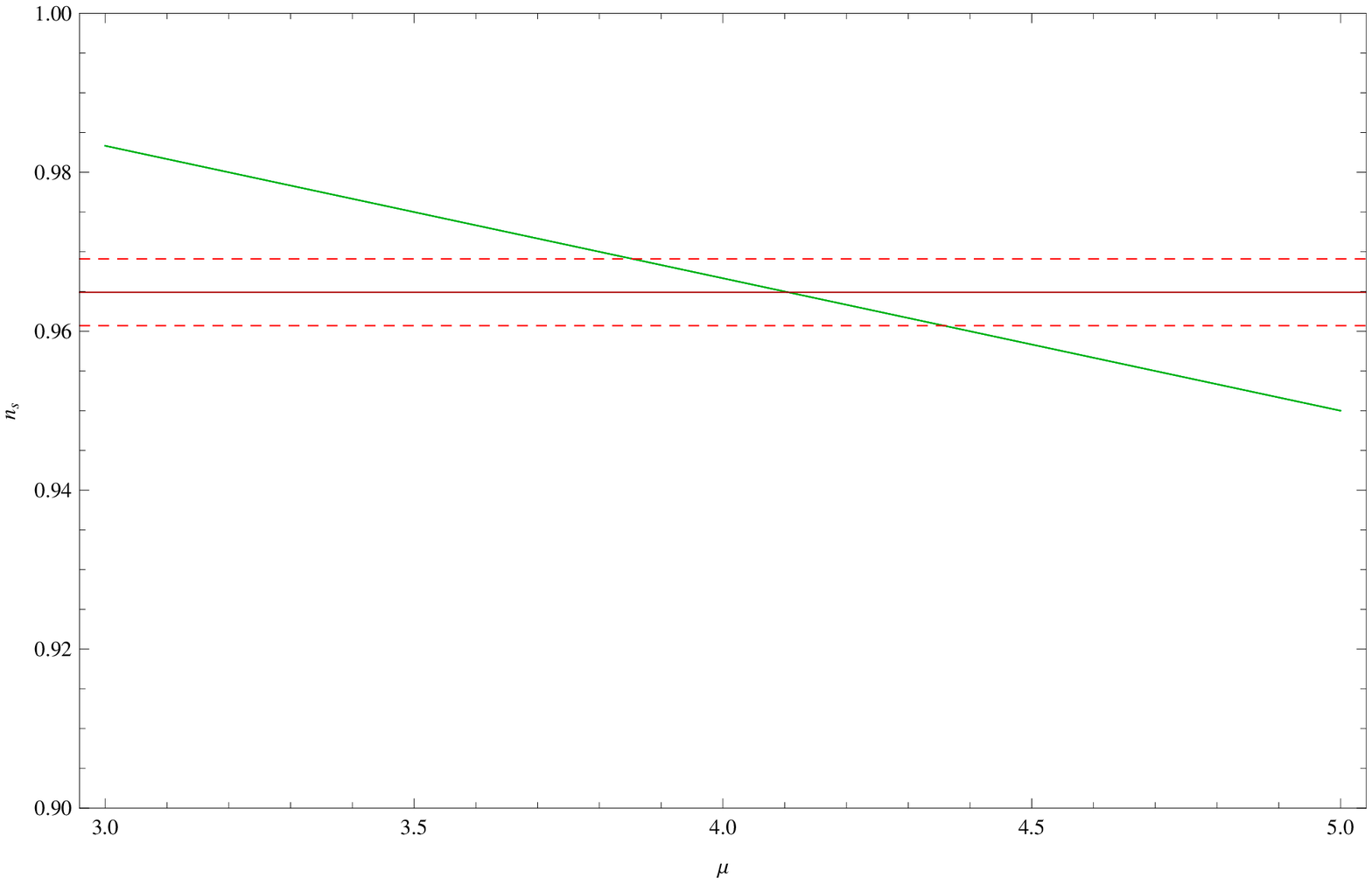}
\includegraphics[width=18pc]{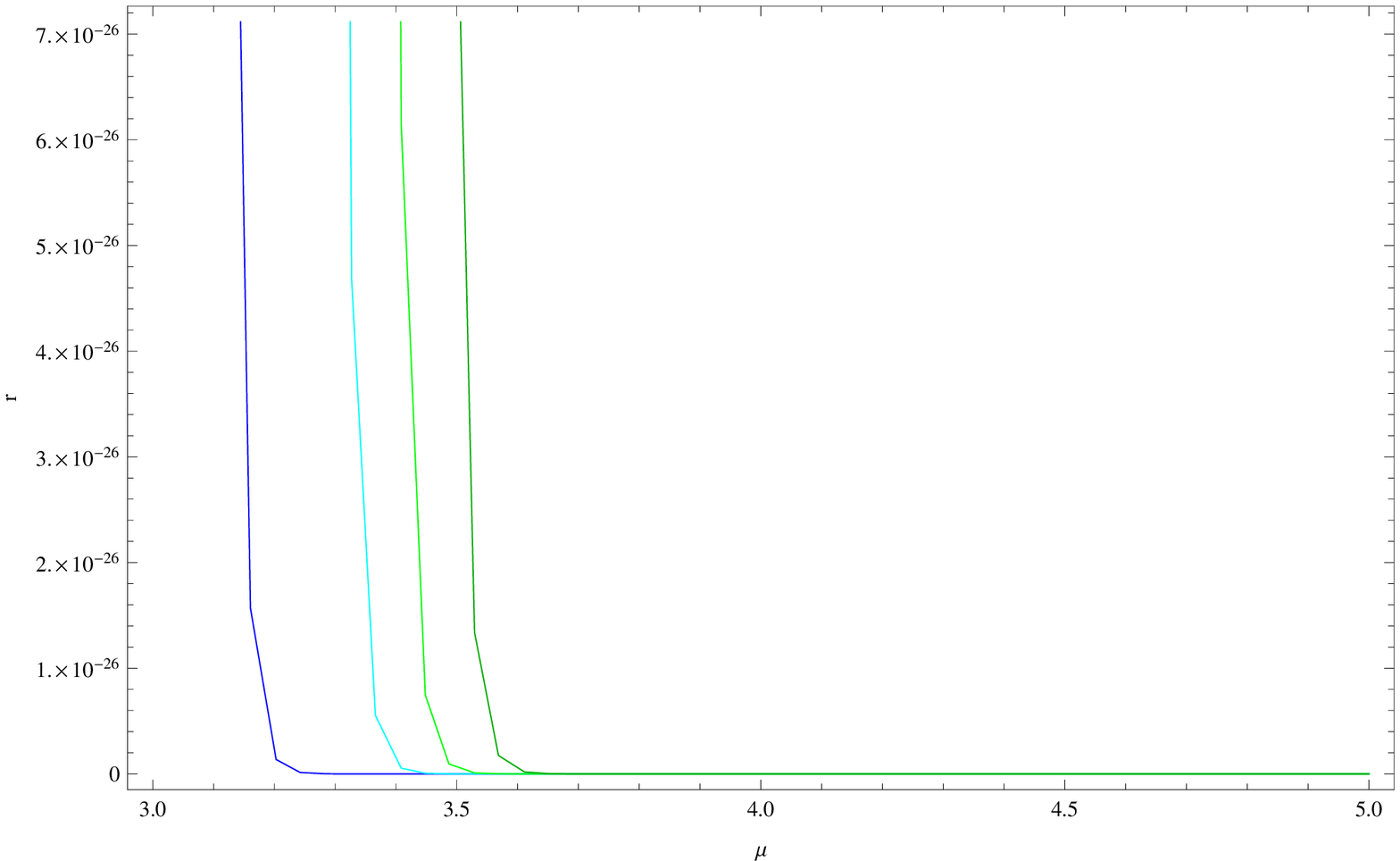}
\caption{The spectral index $n_{S}$ (left plot) and the
tensor-to-scalar ratio $r$ (right plot), for the power-law
function $h(\chi) = \gamma \chi^{b}$ in the case of a de Sitter
evolution, with respect to $b$, for $N=50$ , $\gamma=1$ and
$\mu=10^{12}$sec$^{-1}$. The different colors correspond to
different values of $H_{0}$, varying from
$H_{0}=10^{26}$sec$^{-1}$ (the blue curve) to
$H_{0}=10^{29}$sec$^{-1}$ (the darker green curve). The horizontal
dark red line stands for $n_{S}=0.9649$, while the horizontal
dashed red lines for the limits of its confidence interval,
according to Planck 2018 results. The horizontal black line sets
the limit $r=0.064$ from the same results, while the dashed black
an older upper boundary of $r=0.07$ from the BICEP2/Keck-Array.}
\label{fig:desitter-pow}
\end{figure}

Let us assume that the coupling function is a simple power law,
\begin{equation}
\label{powerlawcoupling} h(\chi) = \gamma \chi^{b} \, ,
\end{equation}
where $\gamma$ and $b$ are real constants, to be used as free
parameters later. Using Eqs.~(\ref{desitter-scalar}) and
(\ref{desitter-time}), we can write the coupling function first as
function of time,
\begin{equation}
h(t) = \gamma \left( \mu^{2} t \right)^{b } \, ,
\end{equation}
and then as a function of the $e$-foldings number,
\begin{equation}
\label{desitter-pow-coupling} h(N) = \gamma \left(
\dfrac{\mu^{2}}{H_{0}} N \right)^{b } \, .
\end{equation}
Using Eq.~(\ref{FRGFRW8}), we may derive the potential as a
function of the $e$-foldings number,
\begin{equation}
\label{desitter-pow-potential} \tilde{V} (N) = \dfrac{8 \gamma
(b-1) b H_{0}^4 \left( \frac{\mu^2}{H_{0}} N \right)^b}{N^2} \, ,
\end{equation}
as well as the Lagrange multiplier,
\begin{equation}
\label{desitter-pow-mult} \lambda (N) = 8 \gamma (b-1) b H_{0}^2
\left( \frac{\mu^2}{H_{0}} N \right)^{b-2} \, .
\end{equation}
 From the equations in (\ref{qfunctionsefold}), we can write the $Q_{i}$
functions with respect to the $e$-foldings number, as follows,
\begin{align}
\label{desitter-pow-qfunctions} Q_{a} (N) &= \dfrac{8\gamma b
H_{0}^3 \left( \frac{\mu ^2}{H_{0}} N \right)^b}{N} \, , \quad
Q_{b} (N) = \dfrac{16\gamma b H_{0}^2 \left( \frac{\mu ^2}{H_{0}}
N \right)^b}{N} \, , \quad
Q_{c} (N) = Q_{d} (N) = Q_{e} (N) = 0 \, , \nonumber \\
Q_{f} (N) &= \dfrac{16\gamma b H_{0}^2 (N + 1 - b) \left(
\frac{\mu^2}{H_{0}} N \right)^b}{N^2} \, , \quad Q_{t} (N) = 1 +
\dfrac{8\gamma b H_{0}^2 \left( \frac{\mu ^2}{H_{0}} N
\right)^b}{N} \, ,
\end{align}
while the wave-speeds appearing in Eqs.~(\ref{wavespeed1}) and
(\ref{wavespeedefold}) are,
\begin{equation}
\label{desitter-pow-wave} c_{A}^{2} = 1\, , \quad c_{T}^{2} =
\frac{8\gamma (b-1) b H_{0}^2 \left( \frac{\mu ^2}{H_{0}} N
\right)^b + N^2}{ 8\gamma b H_{0}^2 N \left( \frac{\mu ^2}{H_{0}}
N \right)^b + N^2 } \, .
\end{equation}
The function $E(R,\chi)$ is written with respect to the
$e$-foldings number in the following way,
\begin{equation}
\label{desitter-pow-E} E(N) = \dfrac{96 a^2 b^2 H_{0}^4 \left(
\frac{\mu ^2}{H_{0}} N \right)^{2 b-2}}{1 + 8\gamma b H_{0} \mu^2
\left( \frac{\mu^2}{H_{0}} N \right)^{b-1}} - 8\gamma (b-1) b
H_{0}^2 \left( \frac{\mu ^2}{H_{0}} N \right)^{b-2} \, .
\end{equation}
Using Eqs.~(\ref{slowrollefold}), (\ref{desitter-hubble}),
(\ref{desitter-pow-coupling}) and (\ref{desitter-pow-E}), we
obtain the slow-roll parameters of the de Sitter evolution case,
which are,
\begin{align} \label{desitter-pow-slowroll}
\epsilon_{1} (N) &= \epsilon_{2} (N) = \epsilon_{3} (N) = 0 \, , \nonumber \\
\epsilon_{4} (N) &= \dfrac{-\frac{3072 a^3 (b-1) b^3 H_{0}^4 \mu^4
\left( \frac{\mu^2}{H_{0}} N \right)^{3 b-4}}{\left( 16\gamma b
H_{0} \mu^2 \left( \frac{\mu ^2}{H_{0}} N \right)^{b-1} + 2
\right)^2}+\frac{192 a^2 b^2 (2 b-2) H_{0}^3 \mu ^2
\left(\frac{\mu ^2 N}{H_{0}}\right)^{2 b-3}}{16\gamma b H_{0}
\mu^2 \left(\frac{\mu ^2 N}{H_{0}}\right)^{b-1}+2} - 8\gamma (b-2)
(b-1) b H_{0} \mu ^2 \left( \frac{\mu ^2}{H_{0}} N
\right)^{b-3}}{2 \left( \frac{192 a^2 b^2 H_{0}^4 \left(
\frac{\mu^2}{H_{0}} N \right)^{2 b-2}}{16\gamma b H_{0} \mu ^2
\left( \frac{\mu ^2}{H_{0}} N \right)^{b-1}+2}-8\gamma (b-1) b
H_{0}^2
\left( \frac{\mu ^2}{H_{0}} N \right)^{b-2}\right)} \, , \nonumber \\
\epsilon_{5} (N) &= \frac{4\gamma b H_{0} \mu ^2 \left(
\frac{\mu^2}{H_{0}} N \right)^{b-1}}{1 + 8\gamma b H_{0} \mu ^2
\left( \frac{\mu^2}{H_{0}} N \right)^{b-1}} \, , \quad
\epsilon_{6} (N) = \frac{8\gamma (b-1) b \mu ^4 \left( \frac{\mu
^2}{H_{0}} N \right)^{b-2}}{1 + 8\gamma b H_{0} \mu ^2 \left(
\frac{\mu ^2}{H_{0}} N \right)^{b-1}} \, .
\end{align}
Using the above results, we can proceed in calculating the
spectral indices, from Eqs.~(\ref{spectralind}),
\begin{align}
\label{desitter-pow-spectral} n_{S} =& 1 + (b-1) \left[64 a^2
(b-2) b^2 H_{0}^4 \left( \frac{\mu ^2}{H_{0}} N \right)^{2 b}+N^2
\left( b - 2 - 24\gamma b H_{0}^2 \left( \frac{\mu ^2}{H_{0}} N
\right)^b
\right) \right. \nonumber \\
& \left. -16\gamma b H_{0}^2 N \left( \frac{\mu ^2}{H_{0}} N
\right)^b \left(b \left(6\gamma H_{0}^2 \left( \frac{\mu
^2}{H_{0}} N \right)^b-1 \right) + 2 \right)\right] \nonumber \\
& \times \left\{ N \left(8\gamma b H_{0}^2 \left( \frac{\mu
^2}{H_{0}} N \right)^b + N \right) \left[ N \left( b
\left(12\gamma H_{0}^2 \left(\frac{\mu^2}{H_{0}} N \right)^b - 1
\right)+ 1 \right) - 8\gamma (b-1) b
H_{0}^2 \left(\frac{\mu ^2}{H_{0}} N \right)^b\right] \right\}^{-1} \, , \nonumber \\
n_{T} =& \frac{16\gamma b (b-1) H_{0}^2 \left( \frac{\mu
^2}{H_{0}} N \right)^b}{N \left( 8\gamma b H_{0}^2 \left(
\frac{\mu ^2}{H_{0}} N \right)^b + N \right)} \, .
\end{align}
and the tensor-to-scalar ratio, from Eq.~(\ref{tensorscalar}),
\begin{equation}
\label{desitter-pow-tensorscalar} r = 16 \left| \frac{a b H_{0}^2
(-b+N+1) \left( \frac{\mu ^2}{H_{0}} N \right)^b}{N \left( 8\gamma
b H_{0}^2 \left( \frac{\mu ^2}{H_{0}} N \right)^b + N \right)
\left( \frac{8\gamma (b-1) b H_{0}^2 \left( \frac{\mu ^2}{H_{0}} N
\right)^b + N^2}{N \left( 8\gamma b H_{0}^2 \left( \frac{\mu
^2}{H_{0}} N \right)^b + N \right)} \right)^{3/2}} \right| \, .
\end{equation}
Having these at hand, we can compare them directly to the Planck
\cite{Akrami:2018odb} and the BICEP2/Keck-Array data
\cite{Array:2015xqh}, which indicate that $n_{S} = 0.9649 \pm
0.0042$ and $r < 0.064$. It can be shown that the viability of the
theory is achieved for a restricted range of values of the free
parameters. Actually, if we set $N=50$ (or $N=60$) to indicate the
end of the inflationary era, it is easy to see that the values of
$H_{0}$, $\gamma$ and $\mu$ do not affect the resulting values. In
effect, we choose $\gamma=1$ and $\mu=1$\,sec$^{-1}$ for
simplicity and $H_{0} =10^{26}$sec$^{-1}$ (or $H_{0} =
10^{27}$sec$^{-1}$). The tensor-to-scalar ratio is constantly
close to zero, while the spectral index coincides with the Planck
data only for $\mu \sim 4$\,sec$^{-1}$. Namely, $n_{S} = 0.9644$
only for $b=3.78$ when $N=50$, or $b=4.136$ for $N=60$ for the
same values, $r \in [10^{-50} , 10^{-20} ]$. In
Fig.~\ref{fig:desitter-pow} we present the plots of the spectral
index and of the tensor-to-scalar ratio as a function of $b$.
As a result, a power-law coupling function for the de Sitter
background evolution, may generate a viable inflationary model,
only under the strict assumption of $h (\chi) \sim \chi^{4}$.

\subsection{An Exponential Coupling Function, $h(\chi) =\gamma \e^{b\chi}$}

\begin{figure}[h!]
\centering
\includegraphics[width=18pc]{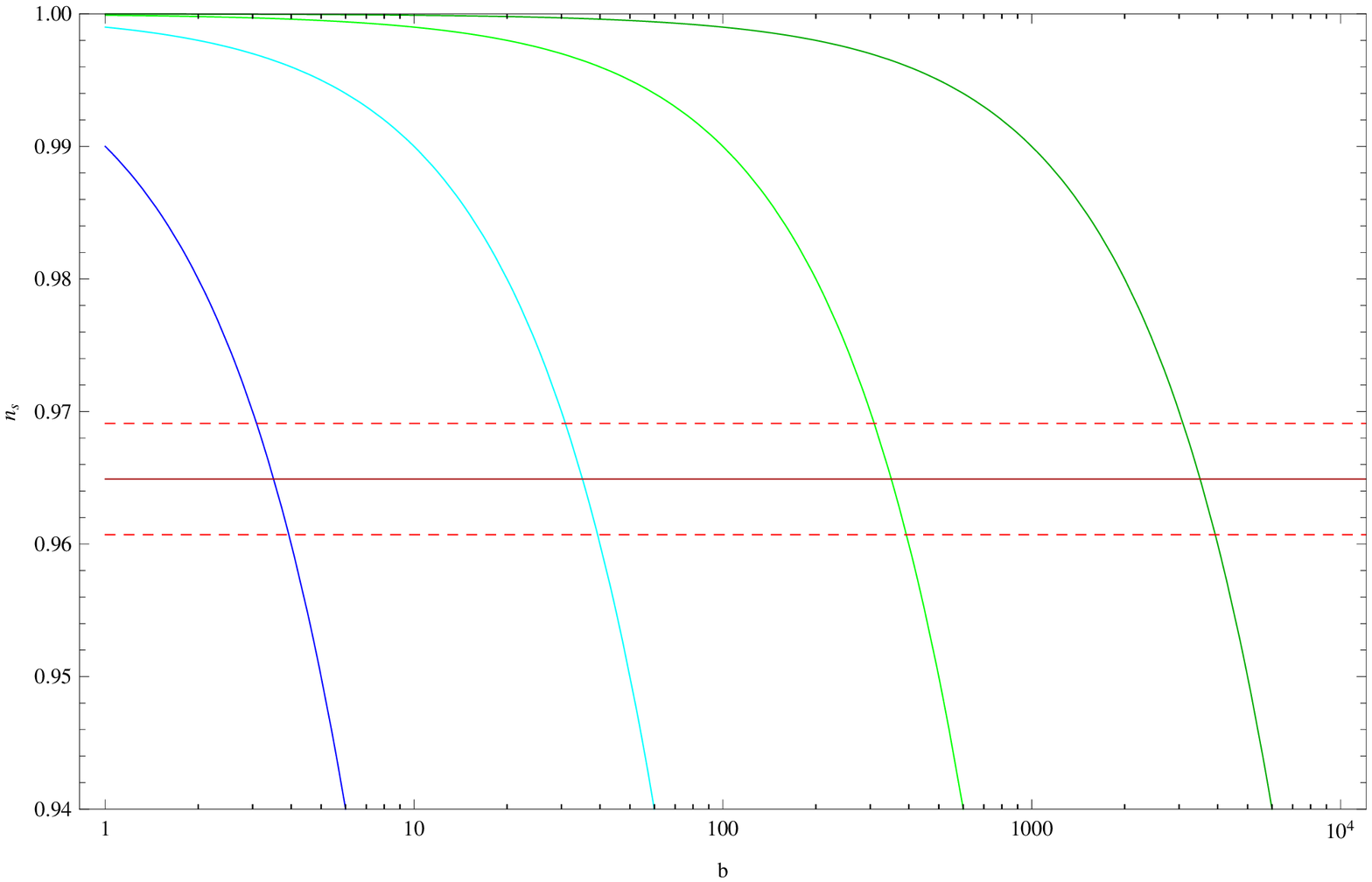}
\includegraphics[width=18pc]{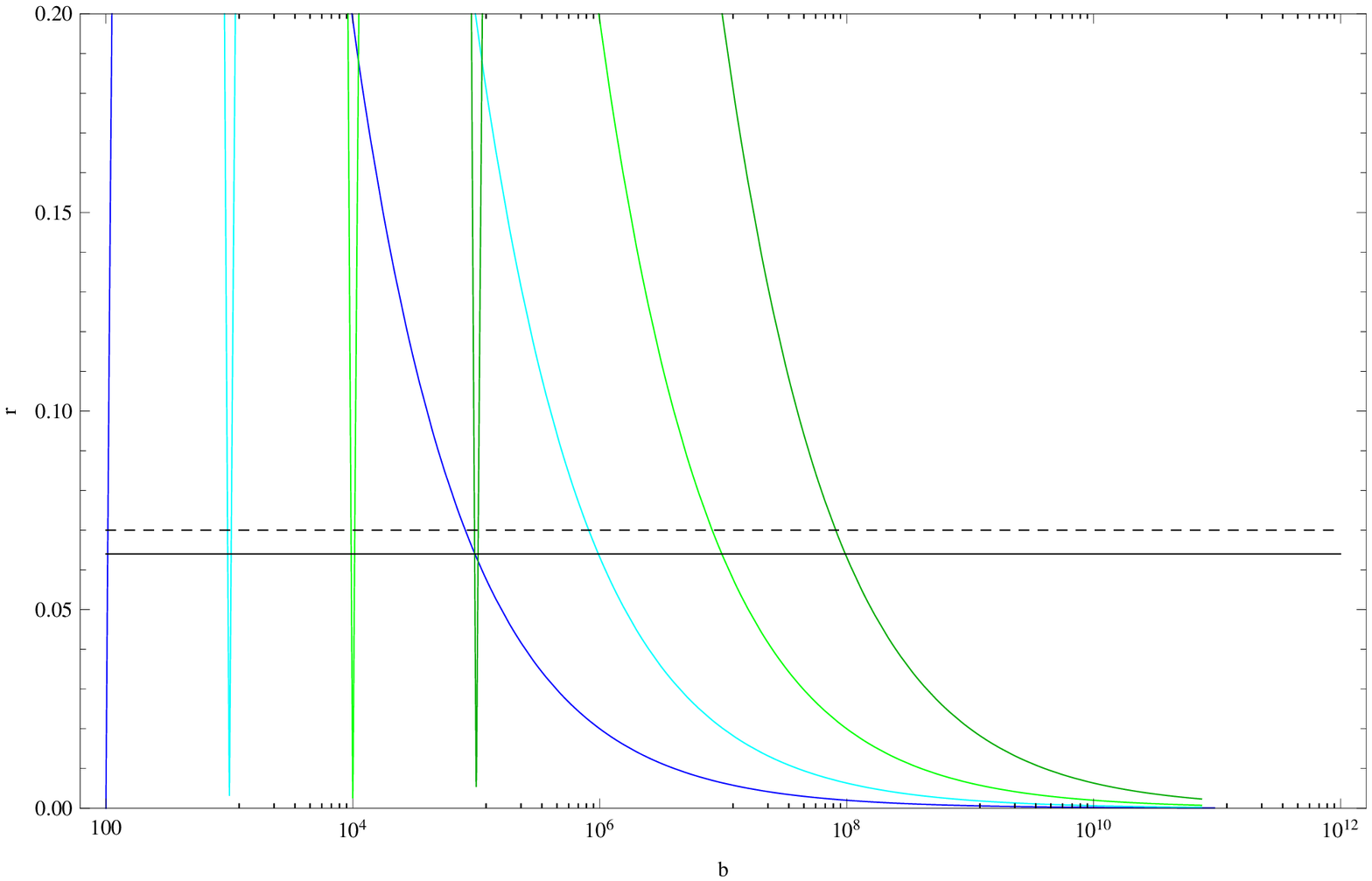}
\caption{The spectral index $n_{S}$ on the left and the
tensor-to-scalar ratio $r$ on the right for the exponential
function $h(\chi)=\gamma \e^{b\chi}$ in the case of a de Sitter
evolution, with respect to $b$, for $N=50$ , $\gamma=1$ and
$\mu=10^{12}$sec$^{-1}$. The color conventions are the same as in
Fig. 1} \label{desitter-b}
\end{figure}

In this case, we assume that the coupling function $h(\chi)$ has
the following exponential form,
\begin{equation}
\label{exponentialcoupling} h(\chi) =\gamma \e^{b \chi} \, ,
\end{equation}
where $\gamma$ and $b$ are real constants, to be used as free
parameters later. Using Eqs.~(\ref{desitter-scalar}) and
(\ref{desitter-time}), we can write the coupling function first as
function of time,
\begin{equation}
h(t) =\gamma \e^{b \mu^{2} t} \, ,
\end{equation}
and then as a function of the $e$-foldings number,
\begin{equation}
\label{desitter-exp-coupling} h(N) =\gamma \e^{\frac{b
\mu^{2}}{H_{0}} N} \, .
\end{equation}
At this point, by using Eq.~(\ref{FRGFRW8}), we may derive the
potential as a function of the $e$-foldings number,
\begin{equation}
\label{desitter-exp-potential} \tilde{V} (N) = 8\gamma b^{2}
H_{0}^{2} \mu^{4} \e^{\frac{b\mu^{2}}{H_{0}} N} \, ,
\end{equation}
as well as the Lagrange multiplier,
\begin{equation}
\label{desitter-exp-mult} \lambda (N) = 8\gamma b^{2} H_{0}^{2}
\e^{\frac{b \mu^{2}}{H_{0}} N} \, .
\end{equation}
Accordingly from Eqs.~(\ref{qfunctionsefold}), we derive the
$Q_{i}$ functions with respect to the $e$-foldings number, as
follows,
\begin{align}
\label{desitter-exp-qfunctions} Q_{a} (N) &= 8\gamma b H_{0}^{2}
\mu^{2} \e^{\frac{b \mu^{2}}{H_{0}} N} \, , \quad Q_{b} (N) =
16\gamma b H_{0} \mu^{2} \e^{\frac{b \mu^{2}}{H_{0}} N} \, , \quad
Q_{c} (N) = Q_{d} (N) = Q_{e} (N) = 0 \, , \nonumber \\
Q_{f} (N) &= 16\gamma b \mu^{2} (H_{0} - b \mu^{2}) \e^{\frac{b
\mu^{2}}{H_{0}} N} \, , \quad Q_{t} (N) = 1 + 8\gamma b H_{0}
\mu^{2} \e^{\frac{b \mu^{2}}{H_{0}} N} \, ,
\end{align}
while the wave-speeds appearing in Eqs.~(\ref{wavespeed1}) and
(\ref{wavespeedefold}), take the following form,
\begin{equation}
\label{desitter-exp-wave} c_{A}^{2} = 1 \, , \quad c_{T}^{2} =
\dfrac{1 + 8\gamma b H_{0}^{2} \mu^{4} \e^{\frac{b \mu^{2}}{H_{0}}
N} }{1 + 8\gamma b H_{0} \mu^{2} \e^{\frac{b \mu^{2}}{H_{0}} N} }
\, .
\end{equation}
The function $E(R,\chi)$ is written with respect to the
$e$-foldings number as follows,
\begin{equation}
\label{desitter-exp-E} E(N) = \dfrac{ 96 a^{2} b^{2} H_{0}^{4}
\e^{\frac{b \mu^{2}}{H_{0}} N} }{ 1 + 8\gamma b H_{0} \mu^{2}
\e^{\frac{b \mu^{2}}{H_{0}} N} } - 8\gamma b H_{0}^{2} \e^{\frac{b
\mu^{2}}{H_{0}} N} \, .
\end{equation}
Using Eqs.~(\ref{slowrollefold}), (\ref{desitter-hubble}),
(\ref{desitter-exp-coupling}) and (\ref{desitter-exp-E}), we
obtain the slow-roll parameters for the de Sitter evolution case
with an exponential coupling function, which is,
\begin{align}
\label{desitter-exp-slowroll}
\epsilon_{1} (N) &= \epsilon_{2} (N) = \epsilon_{3} (N) = 0 \, , \nonumber \\
\epsilon_{4} (N) &= \dfrac{-\frac{3072 a^3 b^4 H_{0}^4 \mu ^4
\e^{\frac{3 b \mu^2}{H_{0}} N}}{\left(16\gamma b H_{0} \mu ^2
\e^{\frac{b \mu^2}{H_{0}} N}+2\right)^2}+\frac{384 a^2 b^3 H_{0}^3
\mu^2 \e^{\frac{2 b \mu ^2}{H_{0}} N}}{16\gamma b H_{0} \mu ^2
\e^{\frac{b \mu^2}{H_{0}} N}+2}-8\gamma b^3 H_{0} \mu ^2
\e^{\frac{b \mu^2}{H_{0}} N}}{2 \left(\frac{192 a^2 b^2 H_{0}^4
\e^{\frac{2 b \mu^2}{H_{0}} N}}{16\gamma b H_{0} \mu ^2
\e^{\frac{b \mu ^2}{H_{0}}
N}+2}-8\gamma b^2 H_{0}^2 \e^{\frac{b \mu ^2}{H_{0}} N}\right)} \, , \nonumber \\
\epsilon_{5} (N) &= \dfrac{4\gamma b H_{0} \mu ^2 \e^{\frac{b \mu
^2}{H_{0}} N}}{1 + 8\gamma b H_{0} \mu ^2 \e^{\frac{b \mu
^2}{H_{0}} N}} \, , \quad \epsilon_{6} (N) = \dfrac{8\gamma b^2
\mu ^4 \e^{\frac{b \mu ^2}{H_{0}} N}}{1 + 8\gamma b H_{0} \mu ^2
\e^{\frac{b \mu ^2}{H_{0}} N}} \, .
\end{align}
By using the above results, we can proceed in calculating the
spectral indices, from Eqs.~(\ref{spectralind}),
\begin{align}
\label{desitter-exp-spectral} n_{S} &= 1 + b \mu ^2
\left(-\frac{1}{8\gamma b H_{0}^2 \mu ^2 \e^{\frac{b \mu ^2
}{H_{0}} N}+H_{0}}+\frac{1}{H_{0}-4\gamma H_{0}^2 \left(3 H_{0}-2
b \mu ^2\right) \e^{\frac{b \mu ^2}{H_{0}}
N}}-\frac{1}{H_{0}}\right) \, , \nonumber \\
n_{T} &= \frac{2 b \mu ^2 \left(\frac{1}{8\gamma b H_{0} \mu ^2
\e^{\frac{b \mu ^2}{H_{0}} N}+1}-1\right)}{H_{0}} \, ,
\end{align}
and the tensor-to-scalar ratio, from Eq.~(\ref{tensorscalar}),
\begin{equation}
\label{desitter-exp-tensorscalar} r = 16 \e^{\frac{b \mu
^2}{H_{0}}  N} \left| \frac{a b \mu ^2 \left(b \mu ^2-H_{0}\right)
\sqrt{8\gamma b \e^{\frac{b \mu^2}{H_{0}} N} H_{0} \mu
^2+1}}{\left( 1 + 8\gamma b^2 \e^{\frac{b \mu^2}{H_{0}} N} \mu ^4
\right)^{3/2}} \right| \, .
\end{equation}
In order to examine the viability of the model, we need to
calculate the numerical values for the spectral index $n_{S}$ and
the tensor-to-scalar ratio $r$, for various values of the
parameters $H_{0}$, $\gamma$, $b$ and $\mu$ at the end of
inflation (for $N \in [ 50 , 60 ]$) and compare these values to
the observational results of the Planck collaboration
\cite{Akrami:2018odb} and the BICEP2/Keck-Array
\cite{Array:2015xqh}. However in this case, no simultaneous
compatibility with the observations can be obtained, and more
specifically, the values of $n_{S}$ and $r$ do not depend on the
choice of $\gamma$, so we set it equal to one for simplicity. They
also do not depend on the number of $e$-foldings, so $N=50$ and
$N=60$ are used in the same manner. They depend on $H_{0}$, $b$
and $\mu$, though, thus assuming that $H_{0} \sim
10^{27}$sec$^{-1}$ and setting $\mu=10^{12}$sec$^{-1}$, we get
$b=35.6$ so that $n_{S} = 0.9644$ (Planck's previous result)
however, the resulting value of the tensor-to-scalar ratio is
excluded. This can also be seen in Fig.~\ref{desitter-b}.


\section{A Flat quasi-de Sitter Vacuum as Background \label{SecV}}

Now we assume that the Universe's evolution is described by the
quasi-de Sitter Hubble rate,
\begin{equation}
\label{quasi-desitter-hubble} H(t) = H_{0} - H_{1} t \, .
\end{equation}
Integrating Eq.~(\ref{quasi-desitter-hubble}) with respect to the
cosmic time, we obtain
\begin{equation*}
N = H_{0} t - \dfrac{H_{1}}{2} t^2 \, ,
\end{equation*}
and solving with respect to time, we may write the latter with
respect to the $e$-foldings number as follows,
\begin{equation}
\label{quasi-desitter-time} t = \dfrac{H_{0} \pm \sqrt{H_{0}^2 - 2
H_{1}  N}}{H_{1}} \, .
\end{equation}
As a result, the Hubble rate with respect to the $e$-foldings
number becomes,
\begin{equation}
H(N) = \pm \sqrt{H_{0}^2-2 H_{1} N} \, ,
\end{equation}
while the Ricci scalar and the Gauss-Bonnet scalar are equal to,
\begin{equation}
\label{desitter-curvatureB} R = 12 \left(H_{0}^2 - 2 H_{1}
N\right) \mp \dfrac{6 H_{1}}{\sqrt{H_{0}^2 - 2 H_{1} N}} \, ,
\quad \mathcal{G} = 24 \left( H_{0}^2 - 2 H_{1} N \right) \left(
H_{0}^2 - H_{1} (2 N+1) \right) \, .
\end{equation}
Finally, we also express the scalar field of Eq.~(\ref{frgdS4})
with respect to the $e$-foldings number as follows,
\begin{equation}
\label{quasi-desitter-scalar} \chi(N) = \mu^{2} \dfrac{H_{0} \pm
\sqrt{H_{0}^2 - 2 H_{1}  N}}{H_{1}} \, .
\end{equation}
As in section \ref{SecIV}, the Eqs.~(\ref{desitter-hubble}),
(\ref{desitter-time}) and (\ref{desitter-scalar}) and a coupling
function allow us to reveal the phenomenological implications of
the model by calculating the observational indices of inflation.

\subsection{An exponential coupling function, $h(\chi) =\gamma \e^{b\chi}$}

At first, we shall assume that the function $h(\chi)$ has the
functional form given in Eq.~(\ref{exponentialcoupling}), which in
the case at hand is written in terms of the $e$-foldings number as
follows,
\begin{equation}
\label{quasi-desitter-exp-coupling} h(N) =\gamma \e^{\frac{b
\left( H_{0} \pm \sqrt{H_{0}^2 - 2 H_{1} N} \right) }{H_{1}}} \, .
\end{equation}
By using Eq.~(\ref{FRGFRW8}), we may derive the potential as a
function of the $e$-foldings number,
\begin{align}
\label{quasi-desitter-exp-potential} \tilde{V} (N) =& 8\gamma b^2
\mu ^4 \left( H_{0}^2 - 2 H_{1} N \right) \exp \left( b \mu ^2
\frac{H_{0} \pm \sqrt{H_{0}^2-2 H_{1} N} }{H_{1}} \pm
\frac{16\gamma b^{2} \mu ^2 H_{1} \left( H_{0}^2-2 H_{1} N \right)
\e^{\frac{b \mu ^2 \left(H_{0} \pm \sqrt{H_{0}^2-2 H_{1} N}
\right)}{H_{1}}}}{\sqrt{H_{0}^2 \pm 2
H_{1} \left( N + \sqrt{H_{0}^2-2 H_{1} N} \right)}} \right) \nonumber \\
& - \frac{2 H_{1} \sqrt{H_{0}^2-2 H_{1} N}}{\kappa ^2
\sqrt{H_{0}^2-2 H_{1} \left(3 \sqrt{H_{0}^2-2 H_{1} N}+N\right)}}
\, ,
\end{align}
as well as the Lagrange multiplier,
\begin{align}
\label{quasi-desitter-exp-mult}
\lambda (N) =&  8\gamma b^2 \left( H_{0}^2-2 H_{1} N \right) \nonumber \\
& \times \exp \left( \pm \left( b \mu^2 \frac{ H_{0} \pm
\sqrt{H_{0}^2-2 H_{1} N} }{H_{1}}-\frac{16\gamma b^{2} H_{1}
\sqrt{(2 H_{1}+1) \left( H_{0}^2-2 H_{1} N \right)} \e^{\frac{b
\mu ^2 \left( H_{0} \pm \sqrt{H_{0}^2-2 H_{1} N}
\right)}{H_{1}}}}{(2 H_{1}+1) \mu ^2}
\right) \right) \nonumber \\
& \pm 2 \frac{H_{1}}{\kappa ^2 \mu^4 \sqrt{H_{0}^2-2 H_{1} N}} \,
.
\end{align}
The functions $Q_{i}$  with respect to the $e$-foldings number are
derived from the Eqs.~(\ref{qfunctionsefold}),
\begin{equation}
\label{quasi-desitter-exp-qfunctions} Q_{a} (N) = Q_{b} (N) =
Q_{c} (N) = Q_{d} (N) = Q_{e} (N) = Q_{f} (N) = 0 \, ,\quad Q_{t}
(N) = 1 \, ,
\end{equation}
while the wave-speeds are
\begin{equation}
\label{quasi-desitter-exp-wave} c_{A}^{2} = 1 \, , \quad c_{T}^{2}
= 1 \, .
\end{equation}
Interestingly, both the $Q_{i}$ functions and the wave-speeds have
a trivial form in the case of the quasi-de Sitter expansion. This
triviality is independent of the coupling function, as we see
later on, and should be attributed to this specific FRW
background.

The function $E(R,\chi)$ with respect to the $e$-foldings number
takes the form,
\begin{align}
\label{quasi-desitter-exp-E} E(N) =& \pm 2 \frac{H_{1}}{\kappa ^2
\mu ^4 \sqrt{H_{0}^2-2 H_{1}
N}} \nonumber \\
& - 8\gamma b^2 \left( H_{0}^2-2 H_{1} N \right) \nonumber \\
& \times \exp \left( b \mu ^2 \frac{ \left( H_{0} \pm
\sqrt{H_{0}^2-2 H_{1} N} \right)}{H_{1}} \pm \frac{16\gamma b^{2}
H_{1} \sqrt{(2 H_{1}+1) \left( H_{0}^2-2 H_{1} N\right)}
\e^{\frac{b \mu ^2 \left( H_{0} \pm\sqrt{H_{0}^2-2 H_{1} N}
\right)}{H_{1}}}}{(2 H_{1}+1) \mu ^2}\right)  \, .
\end{align}
Using Eqs.~(\ref{slowrollefold}), (\ref{desitter-hubble}),
(\ref{desitter-exp-coupling}) and (\ref{desitter-exp-E}), we
obtain the slow-roll parameters of the flat quasi-de Sitter case
with an exponential coupling function. Interestingly, the five of
them take the following trivial form, that seems independent of
the coupling function, while the fourth has a long and complex
form depending on the coupling function,
\begin{equation}
\label{quasi-desitter-exp-slowroll1} \epsilon_{1} (N) =
-\dfrac{H_{1}}{H_{0}^2-2 H_{1} N} \, , \quad \epsilon_{2} (N) =
\epsilon_{3} (N) = 0 \, , \quad \epsilon_{4} (N) =
\varepsilon_\mathrm{exp} (N, H_{0}, H_{1}, a, b, \mu) \, , \quad
\epsilon_{5} (N) = \epsilon_{6} (N) = 0 \, ,
\end{equation}
where $\varepsilon_\mathrm{exp}$ is some notation for the
complicated functional form of the slow-roll index $\epsilon_{4}$.
Similarly, the spectral indices and the tensor-to-scalar ratio are
also long and complex functions of the $e$-foldings number, the
mass $\mu$ and the model parameters, $H_{0}$ and $H_{1}$ due to
the expansion rate and $\gamma$  and $b$ due to the coupling
function, thus we do not present them in close form. What is
interesting to note is that the spectral indices and the
tensor-to-scalar ratio yield the same values independently of
which case of Eq.~(\ref{quasi-desitter-time}) we will use.

\begin{figure}[h!]
\centering
\includegraphics[width=18pc]{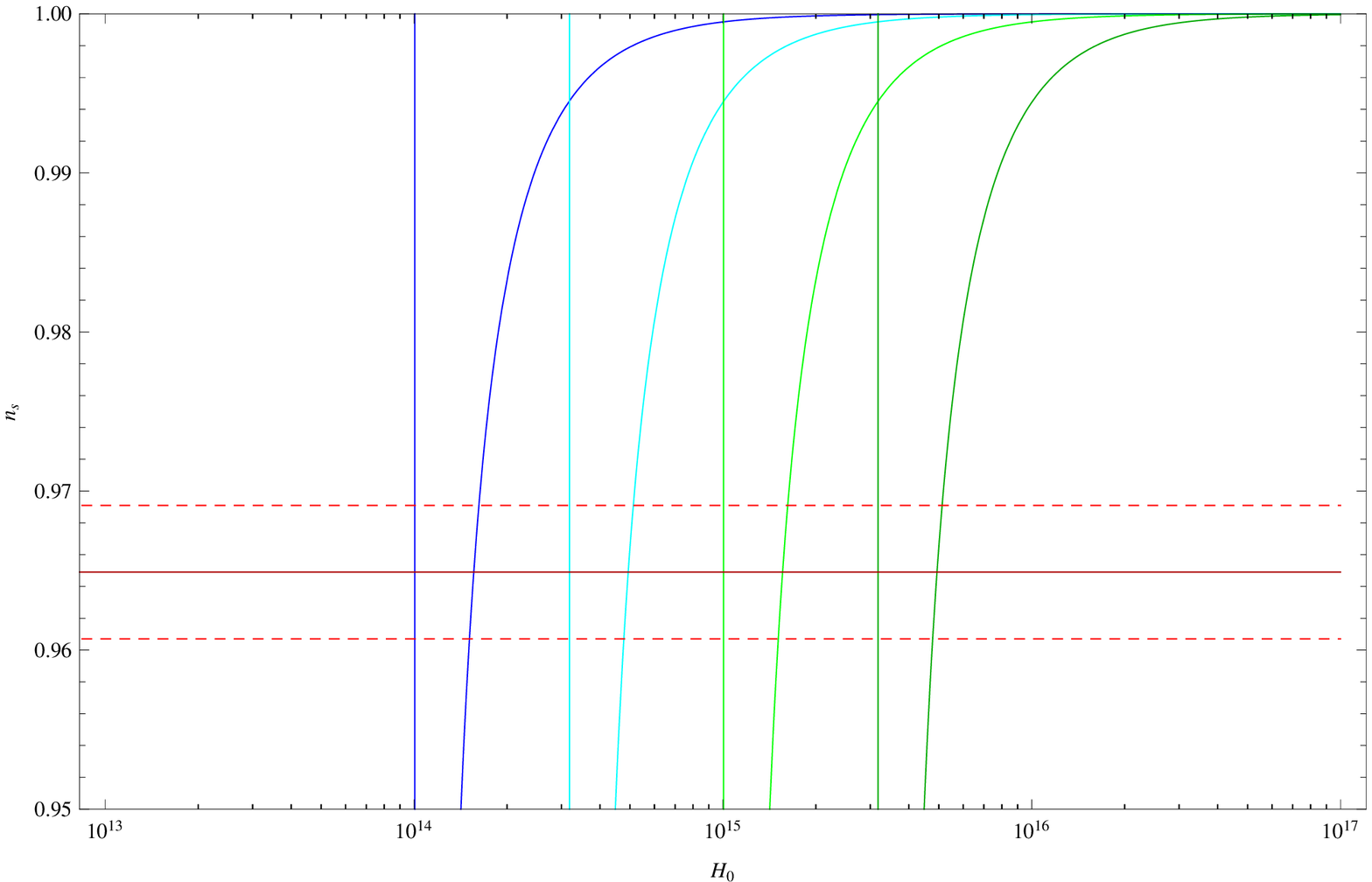}
\includegraphics[width=18pc]{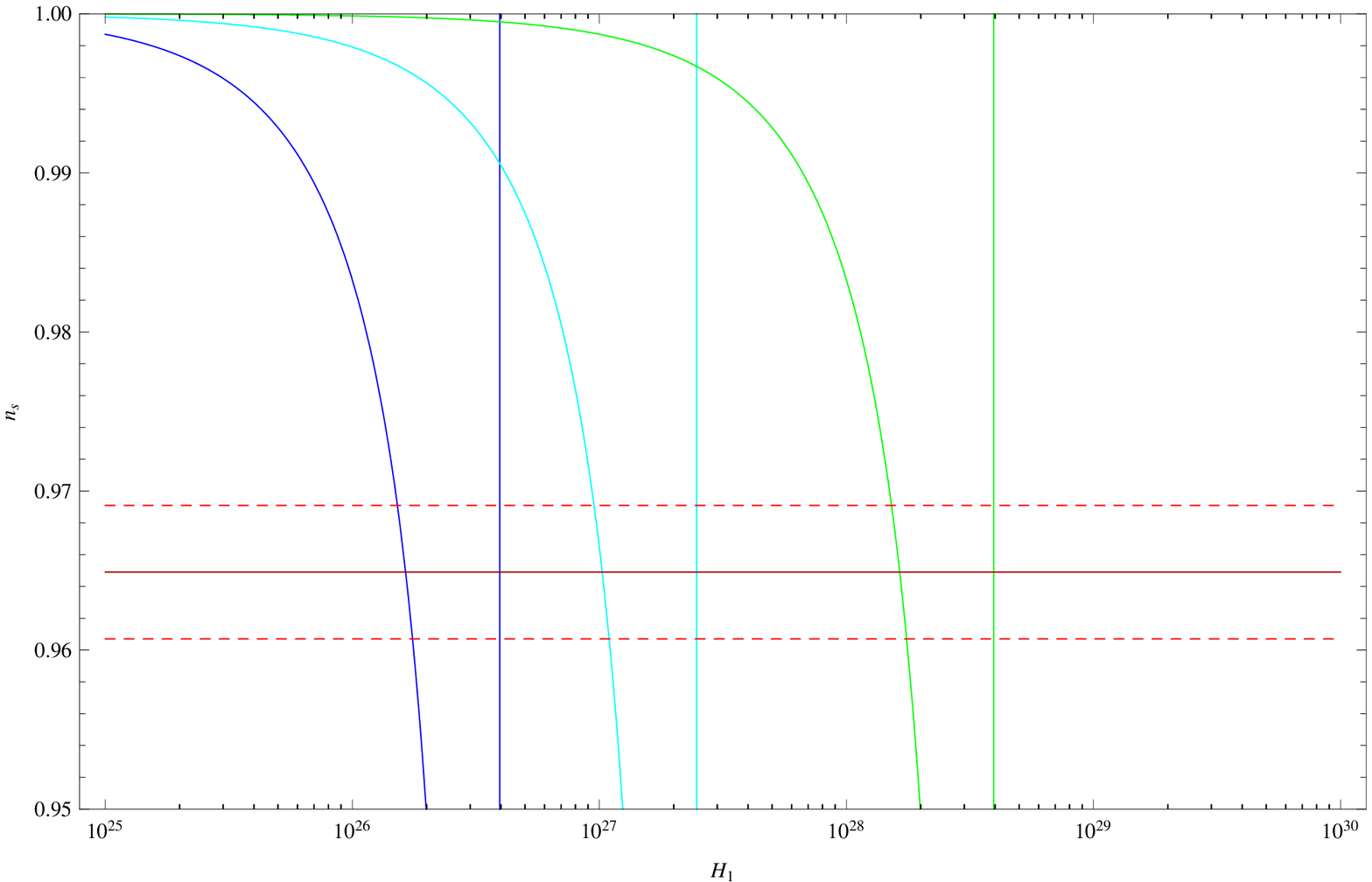}
\caption{The spectral index $n_{S}$ with respect to $H_{0}$ the
left, and to $H_{1}$ in the right, for $N=50$ and $\gamma=b=1$ and
$\mu=1$\,sec$^{-1}$. The plots are identical for $N=60$ and any
other values of $\gamma$ , $b$ and $\mu$. The blue, cyan, green
and darker green curves correspond to different values of $H_{1}$
and $H_{0}$, respectively. The horizontal dark red line stands for
$n_{S}=0.9649$, while the horizontal dashed red lines for the
limits of its confidence interval, according to Planck 2018
results .} \label{fig:quasi-desitter-ns1}
\end{figure}

\begin{figure}[h!]
\centering
\includegraphics[width=18pc]{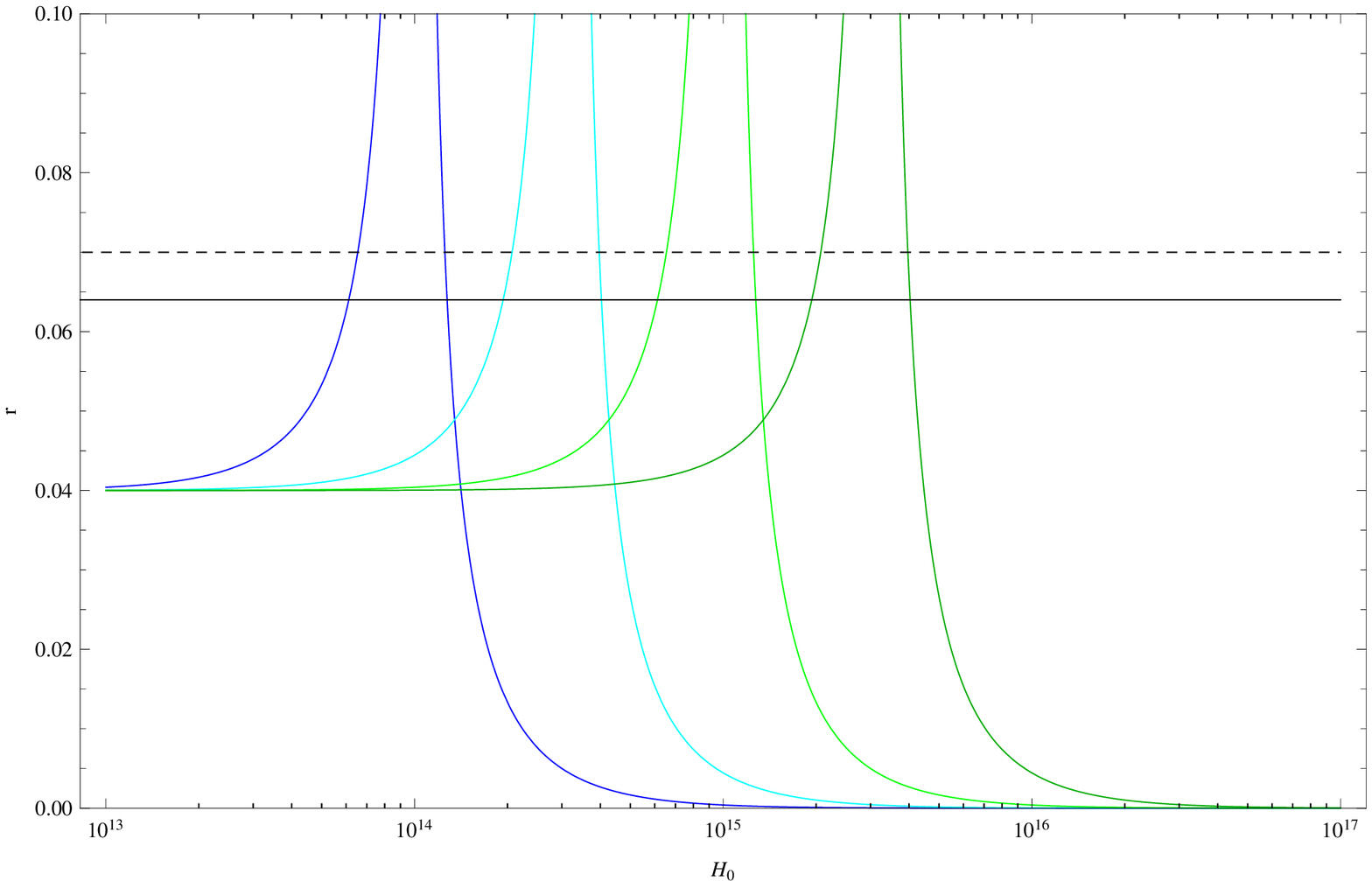}
\includegraphics[width=18pc]{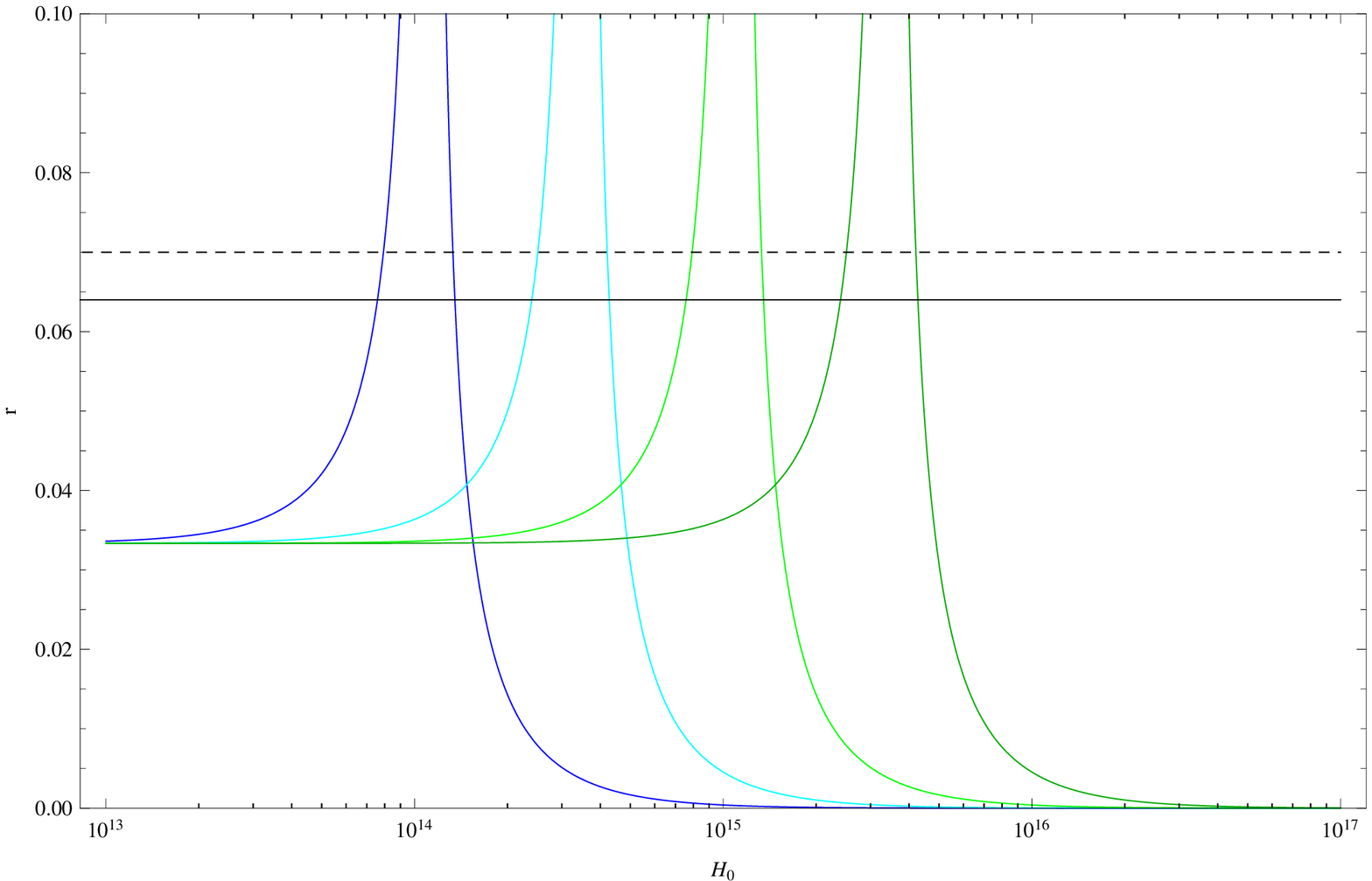}
\caption{The tensor-to-scalar ratio with respect to $H_{0}$ the
left, and to $H_{1}$ in the right, for $N=50$ and $\gamma=b=1$ and
$\mu=1$\,sec$^{-1}$. The plots are identical for $N=60$ and any
other values of $\gamma$ , $b$ and $\mu$. The blue, cyan, green
and darker green curves correspond to different values of $H_{1}$
and $H_{0}$, respectively. The horizontal dashed black line sets
the limit $r<0.07$, while the horizontal black line the limits
$r<0.064$, according to Planck 2015 and Planck 2018 results,
respectively.} \label{fig:quasi-desitter-r1}
\end{figure}

Again, we perform comparisons using the observable values for
$n_{S}$ and $r$ obtained by the Planck  with their latest data
\cite{Akrami:2018odb}, along with \cite{Array:2015xqh}. As we
stated before, the spectral index of the scalar modes must be
within the interval $[0.9607 , 0.9691]$ and mean $n_{S} = 0.9649$;
the tensor-to-scalar mode, on the other hand, is restricted below
$0.1$ by \cite{Array:2015xqh}, while \cite{Akrami:2018odb}
restricts further as $r < 0.064$. In our case, the parameters
$\gamma$ and $b$, as well as the mass $\mu$ of the scalar field
seem not to affect
 the numerical values of the spectral index or the tensor-to-scalar
ratio. As a result, we consider them equal to unity ($\gamma=b=1$
and $\mu=1$\,sec$^{-1}$), so that the analysis is simplified and
focused on the rest of the parameters. The $e$-foldings number is
chosen $N = 50$ and $N=60$, so as to indicate the end of
inflation, but this also does not alter the results. As for the
expansion rate, given that $H_{0} \ge 10^{14}$sec$^{-1}$ for
$H_{1} \approx 10^{26}$sec$^{-2}$ (or that $H_{0} \ge 5 \times
10^{14}$sec$^{-1}$ for $H_{1} \approx 10^{27}$sec$^{-2}$), the
spectral index approaches unity, restricting our choices. We
consider $H_{0}$ to be in the interval $[ 10^{12} ,
10^{15}]\,$sec$^{-1}$ and $H_{1}$ in the respective interval $[
10^{26} , 10^{29} ]\, $sec$^{-2}$, where the spectral index of our
model equals to the observable value, as we can see in
Fig.~\ref{fig:quasi-desitter-ns1}.
For the majority of these cases, the tensor-to-scalar ratio is
close to zero, as we can see in Fig.~\ref{fig:quasi-desitter-r1}.
As an example, choosing $N=50$ (or $N=60$) and $H_{1} =
10^{27}$sec$^{-2}$, then for $H_{0} = 4.91375\times
10^{14}$sec$^{-1}$, we have $n_{S} = 0.9644$ and $r = 0.0282787$,
which comply with  the latest data of the Planck collaboration.
What we need to notice is that these two parameters ($H_{0}$ and
$H_{1}$) need careful fine-tuning and cannot differ significantly
for the set of values we gave, otherwise the model collapses
before the data.

\subsection{A power-law coupling function, $h(\chi) =\gamma \chi^{b}$}

\begin{figure}[h!]
\centering
\includegraphics[width=18pc]{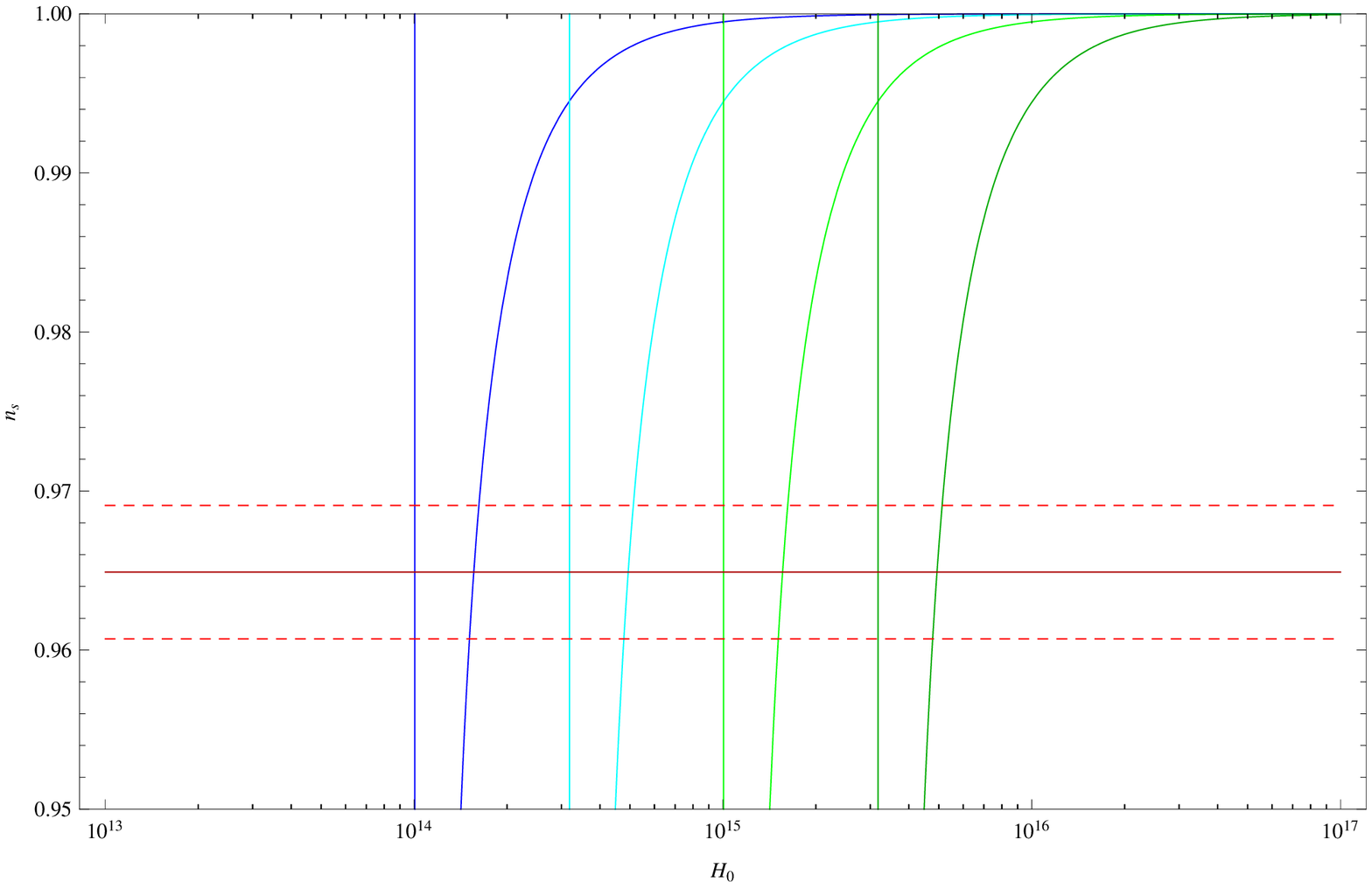}
\includegraphics[width=18pc]{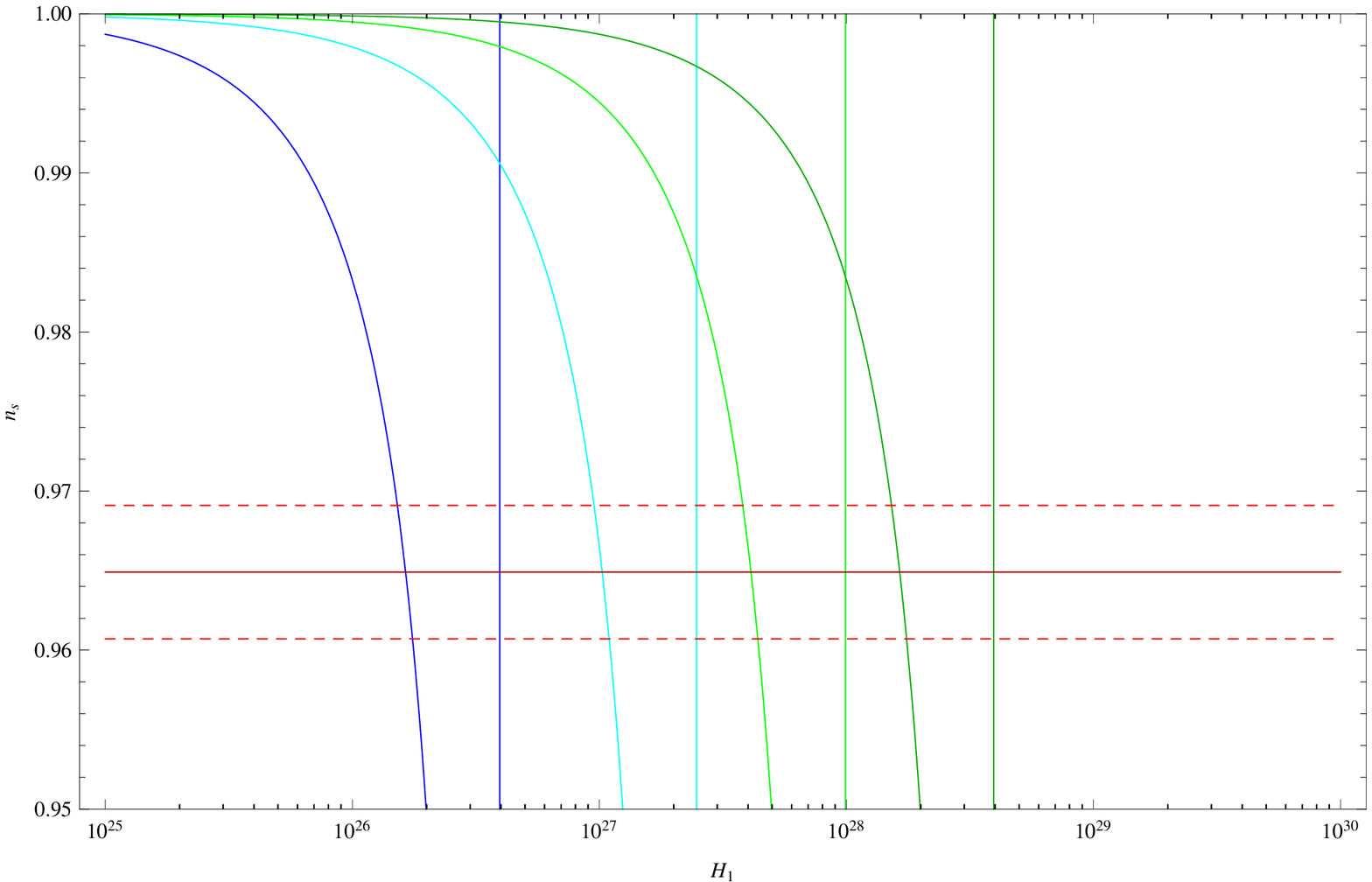}
\caption{The spectral index $n_{S}$ with respect to $H_{0}$ the
left, and to $H_{1}$ in the right, for $N=50$ and $\gamma=b=1$ and
$\mu=1$\,sec$^{-1}$. The plots are identical for $N=60$ and any
other values of $\gamma$ , $b$ and $\mu$. The blue, cyan, green
and darker green curves correspond to different values of $H_{1}$
and $H_{0}$, respectively. The horizontal dark red line stands for
$n_{S}=0.9649$, while the horizontal dashed red lines for the
limits of its confidence interval, according to Planck 2018
results \cite{Akrami:2018odb}.} \label{fig:quasi-desitter-ns2}
\end{figure}

\begin{figure}[h!]
\centering
\includegraphics[width=22pc]{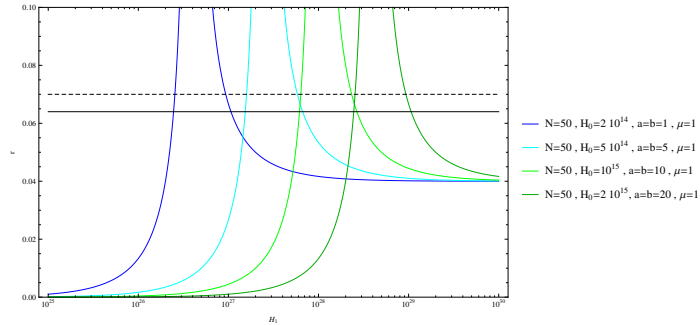}
\caption{The tensor-to-scalar ratio with respect to  $H_{1}$ in
 for $N=50$ and $\gamma=b=1$ and $\mu=1$\,sec$^{-1}$.
The plots are identical for $N=60$ and any other values of
$\gamma$ , $b$ and $\mu$. The blue, cyan, green and darker green
curves correspond to different values of $H_{1}$ and $H_{0}$,
respectively. The horizontal dashed black line sets the limit
$r<0.07$, while the horizontal black line the limits $r<0.064$,
according to Planck 2015 and Planck 2018 results, respectively.}
\label{fig:quasi-desitter-r2}
\end{figure}

Now let us assume that the function $h(\chi)$ takes the form given
in Eq.~(\ref{powerlawcoupling}), which in terms of the
$e$-foldings number is expressed as follows,
\begin{equation}
\label{quasi-desitter-pow-coupling} h(N) = \gamma \e^{\frac{b
\left( H_{0} \pm \sqrt{H_{0}^2 - 2 H_{1} N} \right) }{H_{1}}} \, .
\end{equation}
 From here, using Eq.~(\ref{FRGFRW8}), we may derive the potential
as a function of the $e$-foldings number,
\begin{align}
\label{quasi-desitter-pow-potential}
\tilde{V} (N) =& 8\gamma (b-1) b \mu ^4 \left( H_{0}^2-2 H_{1} N\right) \nonumber \\
& \times \left( \frac{\mu ^2 \left(H_{0} \pm \sqrt{H_{0}^2-2 H_{1}
N}\right)}{H_{1}} \pm \frac{16\gamma b H_{1} \mu ^2
\left(H_{0}^2-2 H_{1} N\right) \left(\frac{\mu ^2 \left(H_{0} \pm
\sqrt{H_{0}^2 \pm 2 H_{1} N}\right)}{H_{1}}\right)^{b-1}}
{\sqrt{H_{0}^2 \pm 2 H_{1} \left(\sqrt{H_{0}^2-2 H_{1}
N}+N\right)}}\right)^{b-2}
\nonumber \\
& - \frac{2 H_{1} \sqrt{H_{0}^2-2 H_{1} N}}{\kappa ^2
\sqrt{H_{0}^2-2 H_{1} \left(3 \sqrt{H_{0}^2-2 H_{1} N}+N\right)}}
\, ,
\end{align}
as well as the Lagrange multiplier,
\begin{align}
\label{quasi-desitter-pow-mult}
\lambda (N) =& \pm 8\gamma (b-1) b \left(H_{0}^2-2 H_{1} N\right) \nonumber \\
& \times \left( \frac{\mu ^2 \left(H_{0} \pm \sqrt{H_{0}^2-2 H_{1}
N}\right)}{H_{1}} \pm \frac{16\gamma b H_{1} \sqrt{(2 H_{1}+1)
\left(H_{0}^2-2 H_{1} N\right)} \left(\frac{\mu ^2 \left(H_{0} \pm
\sqrt{H_{0}^2-2 H_{1} N}\right)}{H_{1}}
\right)^{b-1}}{(2 H_{1}+1) \mu ^2}\right)^{b-2} \nonumber \\
& - \frac{2 H_{1}}{\kappa ^2 \mu ^4 \sqrt{H_{0}^2-2 H_{1} N}} \, .
\end{align}
The $Q_{i}$ functions with respect to the $e$-foldings number have
the same trivial form given in
Eqs.~(\ref{quasi-desitter-exp-qfunctions}) and
(\ref{quasi-desitter-exp-wave}). The function $E(R,\chi)$ with
respect to the $e$-foldings number takes the form,
\begin{align}
\label{quasi-desitter-pow-E} E(N) =& \mp\frac{2 H_{1}}{\kappa ^2
\mu ^4 \sqrt{H_{0}^2-2 H_{1} N}}-8\gamma (b-1) b
\left(H_{0}^2-2 H_{1} N\right) \nonumber \\
& \times \left(\frac{\mu ^2 \left(H_{0} \pm \sqrt{H_{0}^2-2 H_{1}
N}\right)}{H_{1}} \pm \frac{16\gamma b H_{1} \left(H_{0}^2-2 H_{1}
N\right) \left(\frac{\mu ^2 \left(H_{0} \pm \sqrt{H_{0}^2-2 H_{1}
N}\right)}{H_{1}}\right)^{b-1}}{\mu ^2 \sqrt{H_{0}^2-2 H_{1}
\left(-H_{0}^2+2 H_{1} N+N\right)}}\right)^{b-2}  \, .
\end{align}
Using Eqs.~(\ref{slowrollefold}), (\ref{desitter-hubble}),
(\ref{quasi-desitter-pow-coupling}) and
(\ref{quasi-desitter-pow-E}), we obtain the slow-roll parameters
of the flat quasi-de Sitter case with an exponential coupling
function. Except from the fourth one, which has a long and complex
expression,
\begin{equation}
\label{quasi-desitter-pow-slowroll} \epsilon_{4} (N) =
\varepsilon_\mathrm{pow} (N, H_{0}, H_{1}, a, b, \mu) \, ,
\end{equation}
the rest are given in Eqs.~(\ref{quasi-desitter-exp-slowroll1}).
The spectral indices and the tensor-to-scalar ratio have the same
form as in the case of the exponential coupling function,
presented above. Again, we perform comparisons using the
observable values for $n_{S}$ and $r$ obtained by the Planck  with
their latest data \cite{Akrami:2018odb}, along with
\cite{Array:2015xqh}. We assume that the parameters $\gamma$ and
$b$, as well as the mass $\mu$ are equal to unity ($\gamma=b=1$
and $\mu=1$\,sec$^{-1}$), so that the analysis is simplified and
focused on the rest of the parameters. The $e$-foldings number is
chosen $N = 50$ and $N=60$, and as for the expansion rate, given
that $H_{0} \ge 10^{14}$sec$^{-1}$ for $H_{1} \approx
10^{26}$sec$^{-2}$ (or that $H_{0} \ge 5\times 10^{14}$sec$^{-1}$
 for $H_{1} \approx 10^{27}$sec$^{-2}$), the spectral
index approaches unity, restricting our choices. We consider
$H_{0}$ to be in the interval $[ 10^{12} , 10^{15} ]$sec$^{-1}$
and $H_{1}$ in the respective interval $[ 10^{26} , 10^{29}
]$sec$^{-2}$, where the spectral index value of our model becomes
equal to the observable value, as we can see in
Fig.~\ref{fig:quasi-desitter-ns2}.
For the majority of these cases, the tensor-to-scalar ratio is
close to zero, as we can see in Fig.~\ref{fig:quasi-desitter-r2}.
Setting $N=50$ (or $N=60$) and $H_{1} = 10^{27}$sec$^{-2}$, then
for $H_{0} = 8.43822\times 10^{11}$sec$^{-1}$ (or $H_{0} =
10^{12}$sec$^{-1}$), we have $n_{S} = 0.9644$ and $r = 0.0400002$
(or $r = 0.0333335$), that match the latest data of the Planck
collaboration. Again, these two parameters ($H_{0}$ and $H_{1}$)
need careful fine-tuning and cannot differ significantly from the
above values.

\section{The Case of an Exponential Hubble Evolution \label{SecVI}}

Finally, let us assume that the evolution of our Universe is
described by the following Hubble rate,
\begin{equation}
H = H_0 \e^{-\Omega t}\, , \label{hubble1}
\end{equation}
where $H_0$ and $\Omega$ are model parameters with both having
mass dimension [+1]. The Hubble rate of Eq.~(\ref{hubble1})
becomes approximately a quasi de-Sitter like evolution at early
times, when $t\to 0$, that is,
\begin{equation}\label{earlydesitteranalytic}
H\sim H_0-\Omega H_0 t\, ,
\end{equation}
and also the exit from the inflationary epoch occurs at a finite
time $t_f$, which is,
\begin{equation}
t_f = \frac{1}{\Omega} \ln{[H_0/\Omega]}\, . \label{end time1}
\end{equation}
Moreover such exponential type Hubble parameter has been used in
previous works, in the context of $f(R)$ gravity
\cite{Oikonomou:2018npe,Bamba:2014wda} as well as in different
theoretical frameworks \cite{Frampton:2011rh,Liu:2012iba}.
Motivated by such properties of $H = H_0 \e^{-\Omega t}$, here we
use it in the context of ghost free $f(G)$ gravity to describe the
inflationary phase of our Universe we will test the viability of
the model by confronting it with the Planck 2018 constraints.
Also, we can express the cosmic time as a function of the
$e$-foldings number $N$, by using the definition of the latter,
\begin{equation}
N=\int^{t_f}_{t_h} H dt =\frac{H_0}{\Omega} \e^{-\Omega t_h} - 1\,
, \label{$e$-foldings1}
\end{equation}
where $t_h$ is the horizon crossing time instance. Inverting
Eq.~(\ref{$e$-foldings1}), we get $t_h$ in terms of $N$ as
follows,
\begin{equation}
t_h = \chi_h/\mu^2 = \frac{1}{\Omega}
\ln{\left[\frac{H_0}{\Omega(1 + N)}\right]}\, . \label{horizon
croosing time1}
\end{equation}
This expression of $t_h$ is important, since the inflationary
parameters will be calculated at the horizon crossing time
instance. In the following we will calculate the slow-roll indices
and the observational indices of inflation by specifying the
function $h(\chi)$.

\subsection{Exponential coupling : $h(\chi) = \e^{-\alpha \chi}$\label{sub1}}

\begin{figure}[!h]
\begin{center}
\centering
\includegraphics[width=3.5in,height=2.0in]{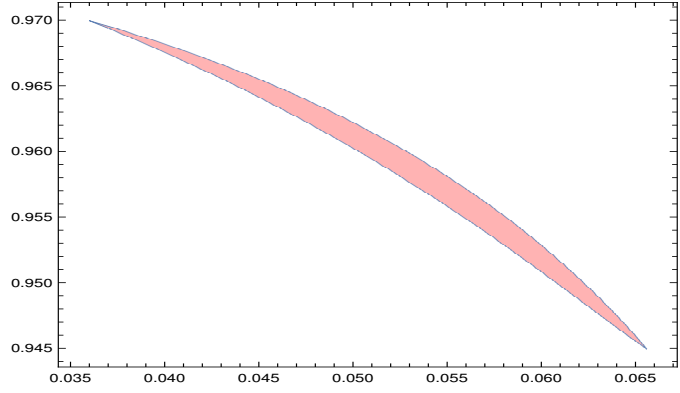}
\caption{Parametric plot of $n_s$ vs $r$ for $0 \lesssim
\Omega/H_0 \leq 0.035$ , $0 \lesssim \alpha\mu^2/\Omega \leq 1.5$
with $\kappa H_0 \sim 0.01$ and $N = 60$.} \label{plot1}
\end{center}
\end{figure}

Let us assume that $h(\chi) = \e^{-\alpha \chi}$ where $\alpha$ is
a model parameter having mass dimension [-1]. For this exponential
function $h(\chi)$, and also for the Hubble rate chosen as in
Eq.~(\ref{hubble1}), the scalar potential is equal to,
\begin{equation}
\tilde{V}(\chi) = \frac{3H_0^2}{\kappa^2} \e^{-2\Omega \chi/\mu^2}
- \frac{2\Omega H_0}{\kappa^2} \e^{-\Omega\chi/\mu^2}
 - 8\alpha H_0^3\mu^2 \e^{-(3\Omega/\mu^2 + \alpha)\chi}\, ,
\label{potential1}
\end{equation}
while the Lagrange multiplier is equal to,
\begin{equation}
\lambda(t) = -\frac{2\Omega H_0}{\kappa^2\mu^4} \e^{-\Omega t} -
\frac{8\alpha H_0^3}{\mu^2} \e^{-(3\Omega + \alpha\mu^2)t}\, .
\label{lambda1}
\end{equation}
Accordingly, the function $E$ defined in Eq.~(\ref{Etime})
evaluated at the horizon crossing time instance, so by expressing
it in terms of the $e$-foldings number, this reads,
\begin{equation}
E(t_h) = \frac{\Omega^2}{\kappa^2} \left[-2(1 + N)
 - 8(1+N)^3 \kappa^2\Omega^2 \left(\frac{\alpha \mu^2}{\Omega}
\right) T^{\alpha \mu^2/\Omega} + 96(1+N)^4 \kappa^4\Omega^4
\left(\frac{\alpha \mu^2}{\Omega}\right)^2 T^{2\alpha
\mu^2/\Omega} \right]\, . \label{E1}
\end{equation}
We also need to evaluate the expression of $\dot{E}$ ($= dE/dt$)
as it will be needed for the calculation of the slow-roll indices.
In terms of the $e$-foldings number, this reads,
\begin{align}
\left. \frac{\dot{E}}{H}\right|_{t_h}=\frac{\Omega^2}{\kappa^2} &
\left[2 + 8(1+N)^2 \kappa^2\Omega^2 \left(\frac{\alpha
\mu^2}{\Omega}\right)^2 T^{\alpha \mu^2/\Omega}
 + 24(1+N)^2 \kappa^2\Omega^2 \left(\frac{\alpha \mu^2}{\Omega}\right)
T^{\alpha \mu^2/\Omega} \right. \nonumber\\
& - \left. 192 (1+N)^3 \kappa^4\Omega^4 \left(\frac{\alpha
\mu^2}{\Omega}\right)^3 T^{2\alpha \mu^2/\Omega}
 - 384 (1+N)^3 \kappa^4\Omega^4 \left(\frac{\alpha \mu^2}{\Omega}\right)^2
T^{2\alpha \mu^2/\Omega}\right] \, , \label{dotE1}
\end{align}
with $T = \frac{\Omega(1 + N)}{H_0}$. Furthermore, by  using
Eq.~(\ref{qfunctionstime}), we explicitly determine the functions
$Q_i$ in terms of $e$-foldings number,
\begin{align}
Q_a(t_h) = & -8 \Omega^3 (1+N)^2 \left[\frac{\alpha\mu^2}{\Omega}
T^{\alpha\mu^2/\Omega}\right] \, , \quad
Q_b(t_h) = -16 \Omega^2 (1+N) \left[\frac{\alpha\mu^2}{\Omega}
T^{\alpha\mu^2/\Omega}\right] \, , \nonumber \\ 
Q_c(t_h) = &  
Q_d(t_h) = 0 \, , \quad 
\left. \frac{Q_e}{H}\right|_{t_h} =16
\Omega^2\left[\frac{\alpha\mu^2}{\Omega}
T^{\alpha\mu^2/\Omega}\right] \, , \nonumber \\ 
Q_f(t_h)=&-8\Omega^2 T^{\alpha\mu^2/\Omega}
\left[\left(\frac{\alpha \mu^2}{\Omega}\right)^2
+ (1+N)\left(\frac{\alpha \mu^2}{\Omega}\right)\right] \, , \quad 
Q_t(t_h)= 1 - 8 \Omega^2 (1+N) \left[\frac{\alpha\mu^2}{\Omega}
T^{\alpha\mu^2/\Omega}\right]\, . \label{Qt1}
\end{align}
Having the above expressions at hand, we can easily calculate the
spectral index $n_s$ and tensor-to-scalar ratio, which are,
\begin{equation}
n_s = 1 - \frac{4}{1 + N}
 - \frac{A_1(\Omega/H_0, \alpha\mu^2/\Omega, \kappa H_0, N)}
{B_1(\Omega/H_0, \alpha\mu^2/\Omega, \kappa H_0, N)}\, ,
\label{spectral index1}
\end{equation}
and
\begin{equation}
r = \left| -\frac{4}{1+N} + 8\kappa^2\Omega^2
T^{\alpha\mu^2/\Omega} \left[\left(\frac{\alpha
\mu^2}{\Omega}\right)^2 + (3 + N)\left(\frac{\alpha \mu^2}
{\Omega}\right)\right] \right|\, , \label{ratio1}
\end{equation}
where $A_1$ and $B_1$ are defined as follows,
\begin{align}
 A_1(\Omega/H_0, & \alpha\mu^2/\Omega, \kappa H_0, N) \nonumber \\
= & \left[2 + 8(1+N)^2 \left(\frac{\alpha \mu^2}{\Omega}\right)^2
\frac{\Omega^2}{H_0^2} \kappa^2H_0^2 T^{\alpha \mu^2/\Omega}
 + 24(1+N)^2 \left(\frac{\alpha \mu^2}{\Omega}\right)
\frac{\Omega^2}{H_0^2} \kappa^2H_0^2 T^{\alpha \mu^2/\Omega} \right. \nonumber\\
& - \left. 192 (1+N)^3 \left(\frac{\alpha \mu^2}{\Omega}\right)^3
\frac{\Omega^4}{H_0^4} \kappa^4H_0^4 T^{2\alpha \mu^2/\Omega}
 - 384 (1+N)^3 \left(\frac{\alpha \mu^2}{\Omega}\right)^2
\frac{\Omega^4}{H_0^4} \kappa^4H_0^4 T^{2\alpha
\mu^2/\Omega}\right]\, , \nonumber
\end{align}
and
\begin{align}
B_1(& \Omega/H_0, \alpha\mu^2/\Omega, \kappa H_0, N) \nonumber \\
& = \left[-2(1 + N) - 8(1+N)^3 \frac{\alpha \mu^2}{\Omega}
\frac{\Omega^2}{H_0^2} \kappa^2H_0^2 T^{\alpha \mu^2/\Omega} 
+ 96(1+N)^4 \left(\frac{\alpha \mu^2}{\Omega}\right)^2
\frac{\Omega^4}{H_0^4} \kappa^4H_0^4 T^{2\alpha
\mu^2/\Omega}\right]\, . \nonumber
\end{align}
It may be noticed that $n_s$ and $r$ depend on the parameters
$\Omega/H_0$, $\alpha\mu^2/\Omega$, $\kappa H_0$ and $N$. We can
now directly confront the spectral index and the tensor-to-scalar
ratio with the Planck 2018 constraints and the BICEP-2 Keck-Array
data, which recall that constraint the observational indices as:
$n_s = 0.9649 \pm 0.0042$ and $r < 0.064$, as shown earlier. For
the model at hand, $n_s$ and $r$ lie within the Planck constraints
for the following ranges of parameter values: $0 \lesssim
\Omega/H_0 \leq 0.035$ , $0 \lesssim \alpha\mu^2/\Omega \leq 1.5$
with $\kappa H_0 \sim 0.01$ and $N = 60$ and this behavior is
depicted in Fig.~\ref{plot1}.
At this stage it deserves mentioning that an exponential coupling
function in a scalar GB theory (without scalar field potential)
admits, at early times, slowly expanding solutions of the form
$a(t) = (At + B)^{1/5}$ (see \cite{Kanti:2015dra}) and thus
exhibits an epoch of deceleration. However here, we show that in
the presence of ghost free $f(\mathcal{G})$ gravity, the
exponential coupling function may be considered as a ``good
inflationary'' model, which allows an early acceleration and also
it is compatible with observations.

Before closing, we can also notice that if some sort of slow-roll
conditions are employed in the model, viability with the
observational data can also be achieved. The slow-roll conditions
in the ghost free Gauss-Bonnet scenario are the following,
\begin{equation}
\left|\dot{H}\right| \ll H^2 \, ,\quad \left|\dot{h}H\right| \ll
1/\kappa^2\, ,\quad \left|\ddot{h}\right| \ll 1/\kappa^2\, .
\label{slow roll conditions}
\end{equation}
The first condition carries the information about the
slow-evolution of the Hubble rate, while the last two demand a
slowly evolving of the function $h(\chi)$. These conditions, and
especially the last two can significantly constrain the parameter
space. For the exponential function $h(\chi)$ we are considering,
the parameters the effectively control the evolution are
$\Omega/H_0$, $\alpha\mu^2/\Omega$, $\kappa H_0$ and $N$. The
slow-roll conditions in the case at hand imply that,
$\frac{\Omega}{H_0} \ll 1$ and
$\left(\frac{\Omega}{H_0}\right)\left(\alpha\mu^2/\Omega\right)
\ll \frac{1}{\kappa H_0}$. Thereby, it is clear that the viable
parametric range i.e $0 \lesssim \Omega/H_0 \leq 0.035$, $0
\lesssim \alpha\mu^2/\Omega \leq 1.5$ with $\kappa H_0 \sim 0.01$
that we considered, is in agreement with the slow-roll conditions.

\subsection{Power Law coupling : $h(\chi) = \left(\frac{\chi}{M}\right)^n$\label{sub2}}

\begin{figure}[!h]
\begin{center}
\centering
\includegraphics[width=3.5in,height=2.5in]{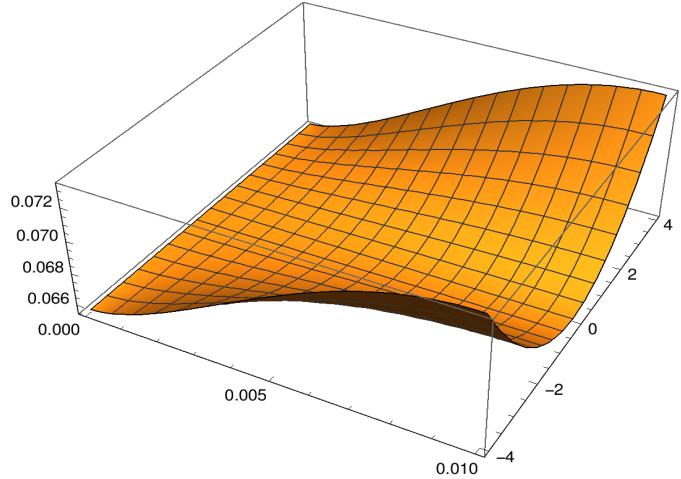}
\caption{3D plot of $r$ vs $\Omega/H_0$ and $\mu^2/(\Omega M)$ for
$\kappa H_0 \sim 0.01$ and $N = 60$.} \label{plot2}
\end{center}
\end{figure}

Let us now assume that the function $h(\chi)$ has the following
form $h(\chi) = \left(\frac{\chi}{M}\right)^n$, where $n$ is a
positive integer and $M$ is a model parameter with mass dimension
[+1]. In this case the scalar potential is,
\begin{equation}
\tilde{V}(\chi) = \frac{3H_0^2}{\kappa^2} \e^{-2\Omega\chi/\mu^2}
 - \frac{2\Omega H_0}{\kappa^2} \e^{-\Omega\chi/\mu^2}
 + \frac{8H_0^3\mu^2}{M^n}n\chi^{n-1} \e^{-3\Omega\chi/\mu^2}\, ,
\label{potential2}
\end{equation}
and the Lagrange multiplier is,
\begin{equation}
\lambda(t) = \frac{8nH_0^3}{\mu^4}\left(\frac{\mu^2}{M}\right)^n
t^{n-1} \e^{-3\Omega t} - \frac{2\Omega H_0}{\kappa^2\mu^4}
\e^{-\Omega t}\, . \label{lambda2}
\end{equation}
Furthermore, the function $E(R)$ and consequently its derivative,
evaluated initially at the horizon crossing time instance, and
expressed eventually in terms of the $e$-foldings number, are
equal to,
\begin{equation}
E(t_h) = \frac{\Omega^2}{\kappa^2} \left[-2(1 + N) - 8n(1+N)^3
\kappa^2\Omega^2 \left(\frac{\mu^2}{\Omega M}\right)^n S^{n-1} +
96 n^2(1+N)^4 \kappa^4\Omega^4 \left(\frac{\mu^2}{\Omega M}
\right)^{2n} S^{2n-2} \right]\, , \label{E2}
\end{equation}
and
\begin{align}
\left. \frac{\dot{E}}{H}\right|_{t_h} = \frac{\Omega^2}{\kappa^2}
& \left[2 + 8 n(n-1)(1+N)^2 \kappa^2\Omega^2
\left(\frac{\mu^2}{\Omega M}\right)^n S^{n-2}
 - 24 n(1+N)^2 \kappa^2\Omega^2 \left(\frac{\mu^2}{\Omega M}\right)^n S^{n-1}
\right. \nonumber\\
& + \left. 192 n^2(1+N)^3 \kappa^4\Omega^4
\left(\frac{\mu^2}{\Omega M}\right)^{2n} S^{2n-3}
 - 384 n^2(1+N)^3 \kappa^4\Omega^4 \left(\frac{\mu^2}{\Omega M}\right)^{2n}
S^{2n-2}\right]\, , \label{dotE2}
\end{align}
with $S = \ln{\left[\frac{H_0}{\Omega(1+N)}\right]}$. Accordingly,
the functions $Q_i$, in terms of $e$-folding number, are equal to,
\begin{align}
Q_a(t_h)=&8 (1+N)^2 n\left(\frac{\mu^2}{\Omega M}\right)^{n}
\Omega^3 S^{n-1} \, , \quad 
Q_b(t_h)=16 (1+N) n\left(\frac{\mu^2}{\Omega M}\right)^{n}
\Omega^2 S^{n-1} \, ,
\nonumber \\
Q_c(t_h)= & 
Q_d(t_h)= 0 \, , \quad 
\left. \frac{Q_e}{H}\right|_{t_h} = -16 n\left(\frac{\mu^2}{\Omega
M}\right)^{n}
\Omega^2 S^{n-1} \, , \nonumber \\ 
Q_f(t_h)=&-8\Omega^2 \left(\frac{\mu^2}{\Omega M}\right)^{n}
\left[n(n-1)S^{n-2} - n(1+N)S^{n-1}\right] \, , \quad 
Q_t(t_h)=1 + 8 (1+N) n\left(\frac{\mu^2}{\Omega M}\right)^{n}
\Omega^2 S^{n-1} \, .\label{Qt2}
\end{align}
Hence, the spectral index becomes in this case,
\begin{equation}
n_s = 1 - \frac{4}{1 + N} - \frac{A_2(\Omega/H_0, \mu^2/(\Omega
M), n, \kappa H_0, N)} {B_2(\Omega/H_0, \mu^2/(\Omega M), n,
\kappa H_0, N)}\, , \label{spectral index2}
\end{equation}
and the tensor-to-scalar ratio is equal to,
\begin{equation}
r = \left| -\frac{4}{1+N} + 8\kappa^2\Omega^2
\left(\frac{\mu^2}{\Omega M}\right)^{n} \left[n(n-1)S^{n-2} - n(3
+ N)S^{n-1}\right] \right|\, , \label{ratio2}
\end{equation}
respectively, with $A_2$ and $B_2$ being defined as follows,
\begin{align}
A_2(\Omega/H_0, &\mu^2/(\Omega M), n, \kappa H_0, N) \nonumber \\
=& \left[2 + 8 n(n-1)(1+N)^2 \left(\frac{\mu^2}{\Omega M}\right)^n
\frac{\Omega^2}{H_0^2} \kappa^2H_0^2 S^{n-2}
 - 24 n(1+N)^2 \left(\frac{\mu^2}{\Omega M}\right)^n \frac{\Omega^2}{H_0^2}
\kappa^2H_0^2 S^{n-1} \right. \nonumber\\
& + \left. 192 n^2(1+N)^3 \left(\frac{\mu^2}{\Omega M}\right)^{2n}
\frac{\Omega^4}{H_0^4} \kappa^4H_0^4 S^{2n-3}
 - 384 n^2(1+N)^3 \left(\frac{\mu^2}{\Omega M}\right)^{2n}
\frac{\Omega^4}{H_0^4} \kappa^4H_0^4 S^{2n-2}\right]\, , \nonumber
\end{align}
and
\begin{align}
B_2(\Omega/H_0, & \mu^2/(\Omega M), n, \kappa H_0, N) \nonumber \\
=&  \left[-2(1 + N) - 8n(1+N)^3 \left(\frac{\mu^2}{\Omega
M}\right)^n
\frac{\Omega^2}{H_0^2} \kappa^2H_0^2 S^{n-1} 
+ 96 n^2(1+N)^4 \left(\frac{\mu^2}{\Omega M}\right)^{2n}
\frac{\Omega^4}{H_0^4} \kappa^4H_0^4 S^{2n-2} \right]\, .
\nonumber
\end{align}
Now we shall confront the resulting theory with the observational
constraints, by assuming two different values for the parameter
$n$, namely, $n=2$ and $n=3$. For $n=2$, the tensor-to-scalar
ratio acquires a minimum value $r_\mathrm{min} =
\left|-\frac{4}{(1+N)}\right|$ which is equal to $r_\mathrm{min} =
0.065$ for $N = 60$ ( and to $r_\mathrm{min}=0.078$ for $N = 50$
). The behavior of the tensor-to-scalar ratio as a function of the
free parameters, is given in Fig.~\ref{plot2}.
As it can be seen in Fig.~\ref{plot2} the minimum value of the
tensor-to-scalar ratio is $r_\mathrm{min} \simeq 0.065$, so the
present model is not viable when the Planck 2018 constraints are
taken into account. For $n = 3$, the theoretical values of $n_s$
and $r$ are found to lie within  the Planck 2018 constraints, when
the values of the free parameters satisfy $0 \lesssim \Omega/H_0
\leq 0.01$, $-4 \lesssim \mu^2/(\Omega M) \leq 1.0$ with $\kappa
H_0 = 0.01$ and $N = 60$.
\begin{figure}[!h]
\begin{center}
\centering
\includegraphics[width=3.5in,height=2.5in]{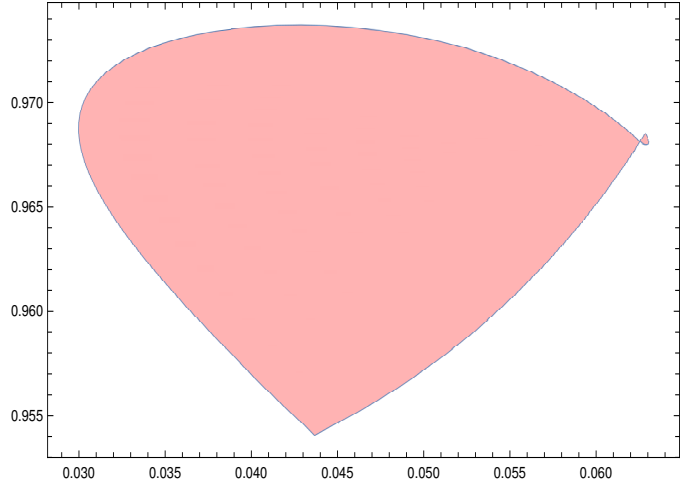}
\caption{Parametric plot of $n_s$ vs $r$ for $0 \lesssim
\Omega/H_0 \leq 0.01$ , $-4 \lesssim \mu^2/(\Omega M) \leq 1.0$
with $\kappa H_0 \sim 0.01$ and $N = 60$.} \label{plot3}
\end{center}
\end{figure}
In Fig.~\ref{plot3} we can see that the spectral index and the
tensor-to-scalar ratio can be simultaneously compatible with the
observational data, for a wide range of values of the free
parameters.

In addition, the cubic coupling function immediately leads to the
slow-roll conditions in terms of the model parameters, which are,
$\frac{\Omega}{H_0} \ll 1$ and
$\left(\frac{\Omega}{H_0}\right)^2\left(\frac{\mu^2}{\Omega
M}\right)^3 \ll \frac{1}{3\kappa^2H_0^2 (1+N)}$, which in turn
indicate that the constraints of values of the free parameters
that lead to a viable phenomenology, which recall are, $0 \lesssim
\Omega/H_0 \leq 0.01$, $-4 \lesssim \mu^2/(\Omega M) \leq 1.0$
with $\kappa H_0 = 0.01$ and $N = 60$, indeed also satisfy the
slow-roll conditions.

Before closing this subsection, we need to comment that it was
shown in \cite{Kanti:2015dra} that a quadratic coupling function
in a scalar GB theory (without scalar field potential) gives
either a pure de-Sitter evolution of our Universe or a de-Sitter
solution at early times connected by a Milne phase at late times,
while the cubic and higher order coupling functions describe
contracting cosmological solutions with a final singularity at
asymptotically infinite time. Thus none of the power law coupling
function corresponding to $n\in [2,3]$ admits a successful
inflationary model in scalar GB theory in the absence of scalar
potential. However in the context of ghost free $f(\mathcal{G})$
gravity, we demonstrated that $h(\chi) \sim \chi^n$ with $n\in
[2,3]$ can realize an accelerating Universe at early times,
although only the cubic coupling function $h(\chi) \sim \chi^3$
produces a viable inflationary phenomenology, in contrast to the
models studied in \cite{Kanti:2015dra}.


\subsection{A Different Reconstruction Approach}\label{sub3}

In this subsection we shall consider an alternative approach in
comparison to the previous cases, by providing the scalar
potential and the Hubble rate, and we seek for the function
$h(\chi )$ that may realize the cosmology with Hubble rate
(\ref{hubble1}). We shall consider two types of potentials, namely
exponential and power law potentials and we shall confront the
resulting theories with the observational data.

\subsubsection{Exponential scalar potential : $\tilde{V}(\chi) = V_0 \e^{-\beta\chi}$}

\begin{figure}[!h]
\begin{center}
\centering
\includegraphics[width=3.5in,height=2.5in]{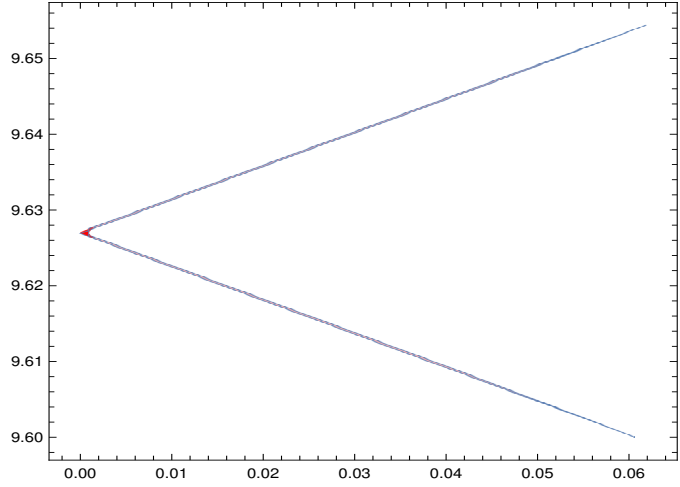}
\caption{Parametric plot of $n_s$ vs $r$ ( $x$ axis $\equiv r$ and
$y$ axis $\equiv 10n_s$ ) for $0.001 \leq \Omega/H_0 \leq 0.02$ ,
$82 \leq \beta\mu^2/\Omega \leq 83$ with $\kappa H_0 \sim 0.01$
and $N = 60$.} \label{plot5}
\end{center}
\end{figure}

Let us first consider an exponential scalar field potential of the
form $\tilde{V}(\chi) = V_0 \e^{-\beta\chi}$, where $V_0$ and
$\beta$ are parameters having mass dimensions [+4] and [-1]
respectively. Using the field equations along with the Hubble
parameter $H = H_0 \e^{-\Omega t}$, one can reconstruct the
coupling function $h(\chi)$, which is in this case,
\begin{align}
h(\chi)=&\frac{1}{8\kappa^2\mu^2H_0^3} \int d\chi \left[2\Omega
H_0 \e^{\Omega\chi/\mu^2} - 3H_0^2 + \kappa^2V_0 \e^{-\beta\chi}
\e^{2\Omega\chi/\mu^2}\right]
\e^{\Omega\chi/\mu^2}\nonumber\\
=&\frac{1}{8\kappa^2H_0^3} \left[H_0 \e^{2\Omega\chi/\mu^2} -
\frac{3H_0^2}{\Omega} \e^{\Omega\chi/\mu^2} +
\left(\frac{\kappa^2V_0}{3\Omega-\beta\mu^2}\right)
\e^{(3\Omega/\mu^2-\beta)\chi}\right]\, , \label{potential3}
\end{align}
and the Lagrange multiplier is,
\begin{equation}
\lambda(t) = \frac{1}{\mu^4} \left[V_0 \e^{-\beta\mu^2 t}
 - \frac{3H_0^2}{\kappa^2} \e^{-2\Omega t}\right]\, .
\label{lambda3}
\end{equation}
With the above expressions of $h(\chi)$ and $\lambda$, we get the
function $E(R)$ as well as its derivative, which are,
\begin{align}
E(t_h) =& \frac{\Omega^2}{\kappa^2} \left[\kappa^2\Omega^2
T^{\frac{\beta\mu^2}{\Omega}} - 3(1 + N)^2 +
\frac{3}{2(N+1)^2}\left(\kappa^2\Omega^2
T^{\frac{\beta\mu^2}{\Omega}}
 - (N+1)(3N+1)\right)^2\right]\, ,
\label{E3} \\
\left. \frac{\dot{E}}{H}\right|_{t_h}=&\frac{\Omega^2}{\kappa^2}
\left[6(1+N) - \frac{\kappa^2\Omega^2
T^{\frac{\beta\mu^2}{\Omega}}}{(1+N)}
\left(\frac{\beta\mu^2}{\Omega}\right)
T^{\frac{\beta\mu^2}{\Omega}}
 - \frac{6}{(1+N)^3} \left(\kappa^2\Omega^2 T^{\frac{\beta\mu^2}{\Omega}}
 - (N+1)(3N+1)\right)^2 \right. \nonumber\\
&+ \left. 3\left(\kappa^2\Omega^2 T^{\frac{\beta\mu^2}{\Omega}} -
(N+1)(3N+1)\right) \left(-\frac{\kappa^2\Omega^2
}T^{\frac{\beta\mu^2}{\Omega}}{(1+N)^3}
\left(\frac{\beta\mu^2}{\Omega}\right)
 - \frac{(3N-1)}{(1+N)^2} + \frac{3\kappa^2\Omega^2 T^{\frac{\beta\mu^2}{\Omega}}}
{(1+N)^3} \right)\right]\nonumber\, , \label{dotE3}
\end{align}
respectively, with $T = \frac{\Omega(N+1)}{H_0}$, defined earlier.
In addition, the functions $Q_i$ as functions of the $e$-foldings
number become in this case,
\begin{align}
Q_a(t_h) = &-\Omega (1+N)^2 \left[-\frac{V_0}{\Omega^2(1+N)^3}
T^{\frac{\beta\mu^2}{\Omega}} + \frac{(3N +1)}{\kappa^2(N +
1)^2}\right] \, ,
\nonumber \\
Q_b(t_h) = &-2 (1+N) \left[-\frac{V_0}{\Omega^2(1+N)^3}
T^{\frac{\beta\mu^2}{\Omega}} + \frac{(3N +1)}{\kappa^2(N +
1)^2}\right] \, ,
\nonumber \\
Q_c(t_h) = & 
Q_d(t_h) =0 \, , \quad 
\left. \frac{Q_e}{H}\right|_{t_h} =
2\left[-\frac{V_0}{\Omega^2(1+N)^3} T^{\frac{\beta\mu^2}{\Omega}}
+ \frac{(3N +1)}{\kappa^2(N + 1)^2}\right] \, ,
\nonumber \\
Q_f(t_h) = &\frac{\mu^2V_0\beta}{\Omega^3(1+N)^3}
T^{\frac{\beta\mu^2}{\Omega}} + \frac{V_0(N-2)}{\Omega^2(1+N)^3}
T^{\frac{\beta\mu^2}{\Omega}}
 - \frac{(3N^2 + N + 2)}{\kappa^2(N+1)^2} \, ,
\nonumber \\
Q_t(t_h) =&1 - (1+N) \left[-\frac{V_0}{\Omega^2(1+N)^3}
T^{\frac{\beta\mu^2}{\Omega}} + \frac{(3N +1)}{\kappa^2(N +
1)^2}\right] \, .
\end{align}
Having the above expressions in hand, we determine the explicit
expressions of the spectral index and  of the tensor-to-scalar
ratio, which are,
\begin{align}
n_s =&  1 - \frac{4}{(1 + N)} - \frac{C_1(\Omega/H_0,
\beta\mu^2/\Omega, \kappa H_0, N)} {D_1(\Omega/H_0,
\beta\mu^2/\Omega, \kappa H_0, N)}\, ,
\label{spectral index3} \\
r =& \frac{3N}{(N + 1)} - \frac{1}{(N + 1)^3}\kappa^2\Omega^2
\left(\frac{\beta\mu^2} {\Omega}\right)
T^{\frac{\beta\mu^2}{\Omega}}
 - \frac{N}{(N + 1)^3}\kappa^2\Omega^2 T^{\frac{\beta\mu^2}{\Omega}}\, ,
\label{ratio3}
\end{align}
respectively, where we took $V_0 = \Omega^4$. Moreover $C_1$,
$D_1$ appearing in Eq.~(\ref{spectral index3})) are defined as
follows,
\begin{align}
C_1(\Omega/H_0, & \beta\mu^2/\Omega, \kappa H_0, N) \nonumber \\
=& \left[6(1+N) - \frac{\kappa^2\Omega^2
T^{\frac{\beta\mu^2}{\Omega} }{(1+N)}
\left(\frac{\beta\mu^2}{\Omega}\right)}
 - \frac{6}{(1+N)^3} \left(\kappa^2\Omega^2 T^{\frac{\beta\mu^2}{\Omega}}
 - (N+1)(3N+1)\right)^2 \right. \nonumber\\
& + \left. 3\left(\kappa^2\Omega^2 T^{\frac{\beta\mu^2}{\Omega}} -
(N+1)(3N+1)\right) \left(-\frac{\kappa^2\Omega^2
T^{\frac{\beta\mu^2}{\Omega}}}{(1+N)^3}
\left(\frac{\beta\mu^2}{\Omega}\right)
 - \frac{(3N-1)}{(1+N)^2} + \frac{3\kappa^2\Omega^2
T^{\frac{\beta\mu^2}{\Omega}}}{(1+N)^3}\right)\right]\, ,
\nonumber \\
D_1(\Omega/H_0, & \beta\mu^2/\Omega, \kappa H_0, N) =
\left[\kappa^2\Omega^2 T^{\frac{\beta\mu^2}{\Omega}} - 3(1 + N)^2
+ \frac{3}{2(N+1)^2}\left(\kappa^2\Omega^2
T^{\frac{\beta\mu^2}{\Omega}} - (N+1)(3N+1)\right)^2\right]\, .
 \nonumber
\end{align}
From Eqs.~(\ref{spectral index3}) and (\ref{ratio3}), it easily
follows that the spectral index of scalar perturbation and the
tensor-to-scalar ratio depend on the dimensionless parameters :
$\Omega/H_0$, $\beta\mu^2/\Omega$, $\kappa H_0$ and $N$. These
theoretical expressions of $n_s$ and $r$ should be confronted with
the latest Planck constraints in order to check the viability of
the model. As a consequence, it is found that the compatibility
with the observational data occurs for a narrow range of values of
the free parameters, and particularly for $0.001 \leq \Omega/H_0
\leq 0.02$, $82 \leq \beta\mu^2/\Omega \leq 83$, $\kappa H_0 \sim
0.01$ and $N = 60$. This can also be seen in Fig.~\ref{plot5}
where we present the parametric plot of $n_s$ and $r$.
With regard to the exponential potential, the classical single
scalar theory has no inherent mechanism to trigger the graceful
exit from inflation, since the slow-roll indices are constant and
field-independent. However the ghost free $f(\mathcal{G})$ theory
has the slow-roll index $\epsilon_4$ which is field dependent, and
thus the slow-roll phase ends when this index becomes of order
$\mathcal{O}(1)$. Moreover we have already shown that the model
with $V=V_0 \e^{-\beta\chi}$ in $f(\mathcal{G})$ gravity, is also
in agreement with Planck observational constraints. Hence the
ghost free $f(\mathcal{G})$ gravity can make the exponential
scalar potential a phenomenologically appealing model for
inflation, in contrast to the single scalar canonical exponential
theory.


\subsubsection{Power law scalar potential}\label{sub4}

\begin{figure}[!h]
\begin{center}
\centering
\includegraphics[width=3.5in,height=2.3in]{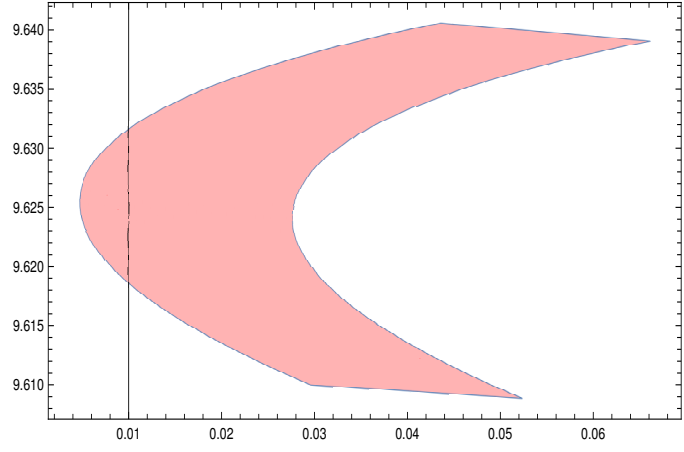}
\caption{Parametric plot of $n_s$ vs $r$ ( $x$ axis $\equiv r$ and
$y$ axis $\equiv 10n_s$ ) for $0.001 \leq \frac{\Omega}{H_0} \leq
0.003$, $50 \lesssim \frac{\mu^2}{\Omega^2}(\kappa H_0)^{2/3}
\lesssim 52$ with $N = 60$.} \label{plot6}
\end{center}
\end{figure}

\begin{figure}[!h]
\begin{center}
\centering
\includegraphics[width=3.5in,height=2.3in]{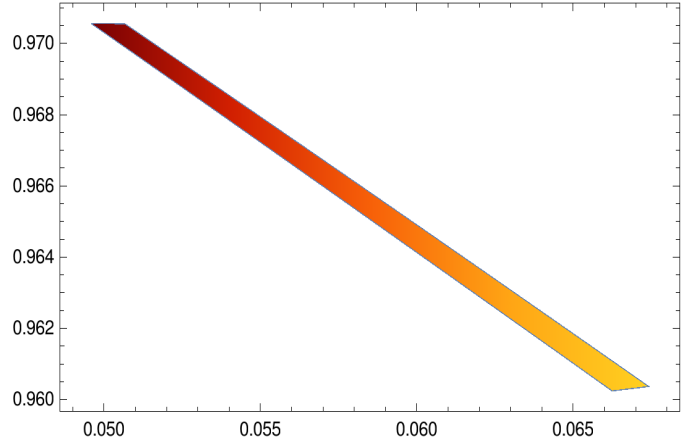}
\caption{Parametric plot of $n_s$ vs $r$ ( $x$ axis $\equiv r$ and
$y$ axis $\equiv n_s$ ) for $0.001 \lesssim \Omega/H_0 \lesssim
0.0026$, $109 \lesssim \frac{\mu^2}{\Omega^2}(\kappa H_0)^{1/2}
\lesssim 110.9$ with $N = 60$.} \label{plot7}
\end{center}
\end{figure}

As a final consideration, we shall assume that the scalar field
potential has the form,
\begin{equation}
\label{scalarpowerlaw} \tilde{V}(\chi) = V_0\chi^{n}\, ,
\end{equation}
where $n$ is a positive integer. For such power law potential, the
function $h(\chi)$ and Lagrange multiplier are equal to,
\begin{align}
h(\chi) =&\frac{1}{8\kappa^2\mu^2H_0^3} \int d\chi \left[2\Omega
H_0 \e^{\Omega\chi/\mu^2} - 3H_0^2 + \kappa^2V_0 \chi^n
\e^{2\Omega\chi/\mu^2}\right]
\e^{\Omega\chi/\mu^2}\nonumber\\
=&\frac{1}{8\kappa^2\Omega H_0^3} \left[ \Omega H_0
\e^{2\Omega\chi/\mu^2} - 3H_0^2 \e^{\Omega\chi/\mu^2} + \left(
\frac{\kappa^2V_0}{3^{1+n}} \chi^n
\left(-\Omega\chi/\mu^2\right)^{-n}\Gamma \left(1+n,
-3\Omega\chi/\mu^2\right) \right) \right]\, , \label{potential4}
\end{align}
and
\begin{equation}
\lambda(t) = \frac{1}{\mu^4} \left[V_0\mu^2 t^n -
\frac{3H_0^2}{\kappa^2} \e^{-2\Omega t}\right]\, , \label{lambda4}
\end{equation}
respectively. Accordingly the function $E(R)$ expressed in terms
of the $e$-foldings number is equal to,
\begin{equation}
E(t_h) = V_0\left(\frac{\mu^2}{\Omega}\right)^n S^n -
\frac{3\Omega^2}{\kappa^2}(N + 1)^2
 + \frac{3\kappa^2}{2\Omega^2(1+N)^2}\left(V_0\left(\frac{\mu^2}{\Omega}\right)^n S^n
 - \frac{\Omega^2}{\kappa^2}(N + 1)(3N + 1)\right)^2\, ,
\label{E4}
\end{equation}
and also its derivative is,
\begin{align}
\left. \frac{\dot{E}}{H}\right|_{t_h} =&
\frac{\mu^2V_0n}{\Omega(1+N)}\left(\frac{\mu^2}
{\Omega}\right)^{n-1} S^{n-1} + \frac{6\Omega^2}{\kappa^2}(1+N) \nonumber \\
& -
\frac{6\kappa^2}{\Omega^2(1+N)^3}\left(V_0\left(\frac{\mu^2}{\Omega}\right)^n
S^n
 - \frac{\Omega^2}{\kappa^2}(N + 1)(3N + 1)\right)^2\nonumber\\
& + 3\kappa^2\left(V_0\left(\frac{\mu^2}{\Omega}\right)^n S^n
 - \frac{\Omega^2}{\kappa^2}(N + 1)(3N + 1)\right) \nonumber \\
& \times \left(\frac{\mu^2V_0n}{\Omega^3(1+N)^3}
\left(\frac{\mu^2}{\Omega}\right)^{n-1} S^{n-1} +
\frac{3V_0}{\Omega^2(1+N)^3} \left(\frac{\mu^2}{\Omega}\right)^n
S^n
 - \frac{(3N - 1)}{\kappa^2(1 + N)^2}\right)\nonumber\, .
\label{dotE4}
\end{align}
For the Hubble rate given in Eq.~(\ref{hubble1}) and with the
expression of $h(\chi)$ we found above, we can easily find the
$Q_i$ functions expressed in terms of the $e$-foldings number,
\begin{eqnarray}
\label{QQ4} Q_a(t_h) &=&-\Omega \left[-\frac{V_0}{\Omega^2(1+N)}
\left(\frac{\mu^2}{\Omega}\right)^n S^n + \frac{(1+3N)}{\kappa^2}\right]\, ,
\nonumber \\
Q_b(t_h) &=&-2\left[-\frac{V_0}{\Omega^2(1+N)^2}
\left(\frac{\mu^2}{\Omega}\right)^n S^n
+ \frac{(1+3N)}{\kappa^2(1+N)}\right] \, , \nonumber \\
Q_c(t_h) &=&0 \, , \nonumber \\ 
Q_d(t_h) &=&0 \, , \nonumber \\ 
\left. \frac{Q_e}{H}\right|_{t_h}
&=&2\left[-\frac{V_0}{\Omega^2(1+N)^3}
\left(\frac{\mu^2}{\Omega}\right)^n S^n +
\frac{(1+3N)}{\kappa^2(1+N)^2}\right] \, ,
\nonumber \\ 
Q_f(t_h) &=&-\frac{n\mu^2V_0}{\Omega^3(1+N)^3}
\left(\frac{\mu^2}{\Omega}\right)^{n-1} S^{n-1} +
\frac{V_0(N-2)}{\Omega^2(1+N)^3}
\left(\frac{\mu^2}{\Omega}\right)^n S^n
 - \frac{(3N^2 + N + 2)}{\kappa^2(1 + N)^2} \, , \nonumber \\ 
Q_t(t_h) &=&1 - \left[-\frac{V_0}{\Omega^2(1+N)^2}
\left(\frac{\mu^2}{\Omega}\right)^n S^n +
\frac{(1+3N)}{\kappa^2(1+N)}\right]\, .
\end{eqnarray}
Let us use the above results in order to investigate the viability
of a power-law class of potentials. According to the latest Planck
data, the cubic and quartic potentials are not compatible with the
Planck data, so let us investigate whether compatibility with the
observations is obtained if the ghost free $f(\mathcal{G})$ theory
is used. Let us first assume that $n=3$ so we consider the cubic
potential first. Using $V(\chi) = V_0\chi^3$ along with the
explicit expressions of $Q_i$ functions (see the equations in
\ref{QQ4}, we determine the spectral index and tensor to scalar
ratio in terms of the model parameters as follows,
\begin{equation}
n_s = 1 - \frac{4}{(1 + N)}
 - \frac{C_2(\Omega/H_0, \frac{\mu^2}{\Omega^2}(\kappa H_0)^{2/3}, N)}
{D_2(\Omega/H_0, \frac{\mu^2}{\Omega^2}(\kappa H_0)^{2/3}, N)}\, ,
\label{spectral index4a}
\end{equation}
and
\begin{equation}
r = \frac{3N}{(N+1)} + \frac{3}{(N+1)^3}
\left(\frac{\mu^2}{\Omega^2} (\kappa H_0)^{2/3}\right)^3
\frac{\Omega}{H_0} S^2
 - \frac{N}{(N+1)^3} \left(\frac{\mu^2}{\Omega^2}(\kappa H_0)^{2/3}\right)^3
\frac{\Omega}{H_0} S^3\, , \label{ratio4a}
\end{equation}
where we assumed that $V_0 = H_0$ (for the cubic potential, $V_0$
has mass dimension [+1]). Moreover $C_2$ and $D_2$ have the
following form,
\begin{align}
C_2& \left(\Omega/H_0, \frac{\mu^2}{\Omega^2}(\kappa H_0)^{2/3}, N
\right)
\nonumber\\
=&\frac{3}{(N+1)} \left(\frac{\mu^2}{\Omega^2}(\kappa
H_0)^{2/3}\right)^3 \frac{\Omega}{H_0} S^2 + 6(N + 1)
 - \frac{6}{(N+1)^3} \left[
\left(\frac{\mu^2}{\Omega^2}(\kappa H_0)^{2/3}\right)^3
\frac{\Omega}{H_0} S^3
 - (N+1)(3N+1) \right]^2\nonumber\\
&+3 \left[\left(\frac{\mu^2}{\Omega^2}(\kappa H_0)^{2/3}\right)^3
\frac{\Omega}{H_0} S^3 - (N+1)(3N+1)\right]
\left[\frac{3}{(N+1)^3}\left(\frac{\mu^2}{\Omega^2}(\kappa
H_0)^{2/3}\right)^3 \frac{\Omega}{H_0} S^2 - \frac{(3N -1)}{(N +
1)^2}\right]\, , \nonumber
\end{align}
and
\begin{align}
D_2 & \left(\Omega/H_0, \frac{\mu^2}{\Omega^2}(\kappa H_0)^{2/3},
N \right)
\nonumber \\
&= \left(\frac{\mu^2}{\Omega^2}(\kappa H_0)^{2/3}\right)^3
\frac{\Omega}{H_0} S^3
 - 3(N+1)^2 
+ \frac{3}{2(N+1)^2} \left[\left(\frac{\mu^2}{\Omega^2}(\kappa
H_0)^{2/3}\right)^3 \frac{\Omega}{H_0} S^3 -
(N+1)(3N+1)\right]^2\, . \nonumber
\end{align}
 It is evident that $n_s$ and $r$ depend on the parameters $\Omega/H_0$,
$\frac{\mu^2}{\Omega^2}(\kappa H_0)^{2/3}$ and $N$. As a result,
it is found that the simultaneous compatibility of $n_s$, $r$ with
Planck 2018 constraints can be achieved for a narrow range of the
free parameters, and in particular for $0.001 \leq
\frac{\Omega}{H_0} \leq 0.003$, $50 \lesssim
\frac{\mu^2}{\Omega^2}(\kappa H_0)^{2/3} \lesssim 52$ and $N =
60$, as shown in Fig.~\ref{plot6}.
We should note that the single canonical scalar field model with
cubic potential without the Gauss-Bonnet coupling yields $n_s
\simeq 0.9089$ and $r\simeq 0.01$, so the spectral index is not
compatible with the Planck data. Hence, the presence of the ghost
free $f(\mathcal{G})$ gravity can make the cubic potential scalar
field class of models to be compatible with the observations. This
kind of result is also shown in a different context
\cite{Odintsov:2018zhw}. Let us now consider the $n=4$ case, in
which case the potential is $V = V_0\chi^4$. In this case, the
spectral index of the primordial scalar curvature perturbations
and the tensor-to-scalar ratio are equal to,
\begin{equation}
n_s = 1 - \frac{4}{(1 + N)}
 - \frac{C_3(\Omega/H_0, \frac{\mu^2}{\Omega^2}(\kappa H_0)^{1/2}, N)}
{D_3(\Omega/H_0, \frac{\mu^2}{\Omega^2}(\kappa H_0)^{1/2}, N)}\, ,
\label{spectral index4b}
\end{equation}
and
\begin{equation}
r = \frac{3N}{(N+1)} + \frac{4}{(N+1)^3}
\left(\frac{\mu^2}{\Omega^2}(\kappa H_0)^{1/2} \right)^4
\frac{\Omega^2}{H_0^2} S^3
 - \frac{N}{(N+1)^3} \left(\frac{\mu^2}{\Omega^2}(\kappa H_0)^{1/2}\right)^4
\frac{\Omega^2}{H_0^2} S^4\, , \label{ratio4b}
\end{equation}
respectively, where $C_3$ and $D_3$ are defined as follows,
\begin{align}
C_3& \left(\Omega/H_0, \frac{\mu^2}{\Omega^2}(\kappa H_0)^{1/2}, N \right)\nonumber\\
=&\frac{4}{(N+1)} \left(\frac{\mu^2}{\Omega^2}(\kappa
H_0)^{1/2}\right)^4 \frac{\Omega^2}{H_0^2} S^3 + 6(N + 1)
 - \frac{6}{(N+1)^3} \left[\left(\frac{\mu^2}{\Omega^2}(\kappa H_0)^{1/2}\right)^4
\frac{\Omega^2}{H_0^2} S^4 - (N+1)(3N+1)\right]^2\nonumber\\
&+ 3 \left[\left(\frac{\mu^2}{\Omega^2}(\kappa H_0)^{1/2}\right)^4
\frac{\Omega^2}{H_0^2} S^4 - (N+1)(3N+1)\right]
\left[\frac{4}{(N+1)^3}\left(\frac{\mu^2}{\Omega^2}(\kappa
H_0)^{1/2}\right)^4 \frac{\Omega^2}{H_0^2} S^3 - \frac{(3N -1)}{(N
+ 1)^2}\right]\, , \nonumber
\end{align}
and
\begin{align}
& D_3 \left(\Omega/H_0, \frac{\mu^2}{\Omega^2}(\kappa H_0)^{1/2},
N \right)
\nonumber \\
&=   \left(\frac{\mu^2}{\Omega^2}(\kappa H_0)^{1/2}\right)^4
\frac{\Omega^2}{H_0^2}
S^4 - 3(N+1)^2 
+ \frac{3}{2(N+1)^2} \left[\left(\frac{\mu^2}{\Omega^2}(\kappa
H_0)^{1/2}\right)^4 \frac{\Omega^2}{H_0^2} S^4 -
(N+1)(3N+1)\right]^2\, . \nonumber
\end{align}
 From Eqs.~(\ref{spectral index4b}) and (\ref{ratio4b}) we can see that $n_s$ and $r$
depend on $\Omega/H_0$, $\frac{\mu^2}{\Omega^2}(\kappa H_0)^{1/2}$
and $N$. In order to examine whether the potential under
consideration provides a viable phenomenology, we need to find the
parametric ranges, if any, for which the theoretical values of
$n_s$ and $r$ match with the latest Planck constraints. A thorough
study of the free parameter values, we found that for $0.001
\lesssim \Omega/H_0 \lesssim 0.0026$, $109 \lesssim
\frac{\mu^2}{\Omega^2}(\kappa H_0)^{1/2} \lesssim 110.9$ and $N =
60$, the inflationary observational indices lie within $0.960 \leq
n_s \leq 0.970$ and $0.049 \leq r \leq 0.065$ respectively. Thus
the model becomes viable (with respect to the Planck 2018
constraints) for such narrow parameter space. However as may be
noticed that $\frac{\mu^2}{\Omega^2}(\kappa H_0)^{1/2}$ must be
fine-tuned within the values $109$ and $110.9$ to keep the model
compatible with Planck constraints. The simultaneous compatibility
of $n_s$ and $r$ is illustrated in Fig.~\ref{plot7}.
However the single canonical scalar field theory with a quartic
potential yields $n_s \simeq 0.8677$ and $r \simeq 0.066$ for 60
$e$-foldings, so the spectral index of the corresponding canonical
scalar field theory is excluded by the latest observational data.
Hence, the presence of ghost free $f(\mathcal{G})$ theory modifies
the quartic scalar field theory and enhances the phenomenological
viability of the model.

\subsection{Stability of First Order Perturbations for the Exponential
Cosmological Evolution}

In this subsection we shall study stability of the first order
perturbation cosmological perturbations, following the work of
\cite{Calcagni:2006ye,Kawai:1997mf,Soda:1998tr,Kawai:1998ab,Sberna:2017nzp,
Sberna:2017xqv} where scalar, vector and tensor perturbations are
calculated in the context of Gauss-Bonnet theory. Scalar, vector
and tensor perturbations are decoupled, as in general relativity,
so that we can focus our attention to tensor and scalar
perturbations separately, as discussed in what follows. Let us
consider first tensor perturbations, which the flat FRW perturbed
line element has the form,
\begin{equation}
ds^2 = -dt^2 + a^2(t) \left[\delta_{ij} + f_{ij}\right]dx^idx^j \,
, \label{tensor per1}
\end{equation}
where $f_{ij}(t,\vec{x})$ is the tensorial perturbation satisfying
$f^i_i = f^{ij}_{,j} = 0$. Plugging back the above metric into the
original action and expanding, keeping terms up to
$\mathcal{O}(f^2)$ (to obtain the first order equations), we get
the following perturbed action
\cite{Calcagni:2006ye,Sberna:2017nzp,Sberna:2017xqv},
\begin{equation}
\delta S_f = \int d^4x a^3(t) \left[\left(1 +
8\kappa^2\dot{h}H\right) \dot{f}_{ij}\dot{f}^{ij}
 - \frac{1}{a^2} \left(1 + 8\kappa^2\ddot{h}\right) f_{ij,k}f^{ij,k}\right] \, ,
\label{tensor per2}
\end{equation}
where we use the background equations of motion. With the Fourier
decomposition as $f_{ij}(t,\vec{x}) = \int dk \tilde{f}_{ij}(t)
\e^{i\vec{k}.\vec{x}}$, the above perturbed action takes the
following form,
\begin{equation}
\delta S_f = \int d\vec{k}dt a^3(t) \left[\left(1 +
8\kappa^2\dot{h}H\right) \dot{\tilde{f}}_{ij}\dot{\tilde{f}}^{ij}
 - \frac{1}{a^2} \left(1 + 8\kappa^2\ddot{h}\right) \tilde{f}_{ij,k}\tilde{f}^{ij,k}\right] \, .
\label{tensor per3}
\end{equation}
Thereby, the tensor perturbation is ghost free and stable if the
following two conditions hold true,
\begin{align}
1 + 8\kappa^2\dot{h}H>&0 \, , \nonumber\\
1 + 8\kappa^2\ddot{h}>&0 \, , \label{condition}
\end{align}
and are satisfied simultaneously. If we assume that the slow-roll
conditions of Eq.~(\ref{slow roll conditions}) hold true, the
coupling function $h(\chi)$ rolls slowly if it obeys
$\left|\dot{h}H\right| \ll 1/\kappa^2$ and $\left|\ddot{h}\right|
\ll 1/\kappa^2$. We have shown in previous sections that a
phenomenologically viable cosmological evolution also satisfies
these constraints if the free parameters are chosen appropriately,
so in view of Eq.~(\ref{condition}) we may conclude that the
tensor perturbations are ghost free and stable, at least at first
order. So the theory is compatible with the observational data and
stable up to first order cosmological tensor perturbations.

\begin{figure}[!h]
\begin{center}
\includegraphics[width=2.5in,height=2.0in]{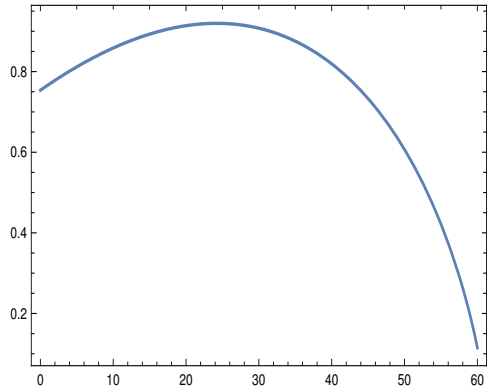}
\caption{$\frac{\dot{h}H}{\ddot{h}}$ vs. $e$-folding number for
$h(\chi)\sim \chi^3$ and with $\frac{\Omega}{H_0}=0.001$ and
$\mu^2/(\Omega M)=0.5$.} \label{plot8}
\end{center}
\end{figure}

Now let us turn our focus to scalar perturbations on the FRW
background spacetime, in which case the line element is,
\begin{equation}
ds^2 = -(1 + 2\Psi)dt^2 +a^2(t) (1 - 2\Psi) \delta_{ij}dx^idx^j\,
, \label{scalar per1}
\end{equation}
with $\Psi(t,\vec{x})$ being the scalar perturbation. Following
\cite{Calcagni:2006ye}, the perturbed action up to order
$\mathcal{O}(\Psi^2)$ is equal to,
\begin{equation}
\delta S_{\Psi} = \frac{1}{2} \int d^4x
a^3(t)Z_1\left[\dot{\Psi}^2 - \frac{Z_2}{a^2} \left(\partial_i
\Psi \right)^2\right]\, , \label{scalar per2}
\end{equation}
where $Z_1$ and $Z_2$ are defined as follows,
\begin{equation}
Z_1 = \frac{-\mu^4\lambda + \frac{3(8\kappa^2\dot{h}H^2)^2}
{2\kappa^2(1+8\kappa^2\dot{h}H)}} {\left(H -
\frac{4\kappa^2\dot{h}H^2}{1+8\kappa^2\dot{h}H}\right)^2} \, ,
\quad Z_2 = 1 + 4 \frac{(\ddot{h} - \dot{h}H)
\left(\frac{8\kappa^2\dot{h}H^2} {1+8\kappa^2\dot{h}H}\right)^2}
{-\mu^4\lambda +
\frac{3(8\kappa^2\dot{h}H^2)^2}{2\kappa^2(1+8\kappa^2\dot{h}H)}}\,
. \label{scalar per3}
\end{equation}
Clearly the scalar perturbation is ghost free and stable if $Z_1$
and $Z_2$ are both positive. With the slow-roll criteria taken
into account, and the corresponding field equations, the
positivity of $Z_1$, $Z_2$ is guaranteed if the following two
conditions hold true,
\begin{equation}
\dot{H} < 0\, , \quad \ddot{h} - \dot{h}H > 0\, . \label{scalar
condition}
\end{equation}
Now let us proceed to explore whether, for our considered choice
of coupling or potential function, the above two conditions are in
agreement with the Planck 2018 constraints. The first condition is
satisfied for $\Omega > 0$ ( as $\dot{H} = -\Omega H_0 \e^{-\Omega
t}$ ) and simply gives the information that the Hubble parameter
must decrease with cosmic time at the early universe, which is
also expected in an inflationary scenario. On the other hand, it
is shown that all the previous four cases (see Sections from
[\ref{sub1}] to [\ref{sub4}]) need $\Omega > 0$ in order to be
compatible with Planck constraints and thus one of the stability
condition of scalar perturbation is ensured. Now let us
investigate the second condition case by case: for the exponential
coupling i.e. $h(\chi) = \e^{-\alpha\chi}$, $\ddot{h} - \dot{h}H$
becomes positive for $\alpha > 0$ which is also needed to make the
model observationally viable (as shown in Section [\ref{sub1}]).
To investigate what happens in the power-law case of $h(\chi)$, we
provide the plot of $\frac{\dot{h}H}{\ddot{h}}$ as a function of
the $e$-foldings number in Fig.~\ref{plot8}.

As it can be seen in Fig.~\ref{plot8}, the ratio
$\frac{\dot{h}H}{\ddot{h}}$ remains less than unity for all the
parameter values that render the theory compatible with the latest
Planck data. Thus this ensures numerically the stability of the
scalar perturbations.

\subsection{Reheating mechanism for the exponential cosmological evolution}

Before moving to the conclusion section, here we discuss the
phenomenological implications of the theory we studied in the
reheating era, and the possible effects of Gauss-Bonnet coupling
on it. Needless to say that reheating describes the production of
Standard Model matter after the period of accelerated expansion.
For this purpose, we assume that the inflaton field (i.e the field
$\chi$) is coupled to another scalar field $\zeta$, given by the
interaction Lagrangian,
\begin{eqnarray}
 L_{int} = -g\lambda \chi\zeta^2\, ,
 \label{reheating1}
\end{eqnarray}
where $g$ is a dimensionless coupling constant and $\lambda$ is a
mass scale. The scalar field $\zeta$ quantifies Standard Model
particles in our case study. With this interaction Lagrangian, the
decay rate of the inflaton into $\zeta$ particles becomes,
\begin{eqnarray}
 \Gamma = \frac{g^2\lambda^2}{8\pi m}\, ,
 \label{reheating2}
\end{eqnarray}
where $m$ denotes the mass of the inflaton field and can be
obtained from the effective potential $V_{eff}(\chi) =
\tilde{V}(\chi) - 24H^2(H^2 + \dot{H})h(\chi)$ through which the
Gauss-Bonnet coupling function ($h(\chi)$) affects indirectly the
reheating mechanism. Moreover, the presence of Gauss-Bonnet term
also affects the self-potential function $\tilde{V}(\chi)$ as may
be noticed in Eq. (\ref{FRGFRW8}) ( see the terms dependent on
$h(\chi)$ in the right hand side of Eq. (\ref{FRGFRW8})).
Generally during the reheating epoch, the inflaton losses energy
due to the expansion of the Universe, and due to transfer of
energy to the $\zeta$ particles, controlled by the Hubble
parameter and the decay rate respectively. As a result the
production of $\zeta$ particles becomes effective when the Hubble
parameter becomes less or comparable to $\Gamma$, otherwise the
energy loss into particles is negligible compared to the energy
loss due to the expansion of space as occurred during the early
phases of the inflation. Therefore, the time scale $t_h$ (let us
call it the reheating time) after when the production of $\zeta$
becomes effective is given by,
\begin{eqnarray}
 H(t_h) = \Gamma = \frac{g^2\lambda^2}{8\pi m}\, .
 \label{reheating4}
\end{eqnarray}
Thus the reheating time depends on the mass of the inflaton field.
For the purpose of determining the inflaton mass explicitly, we
consider two different coupling functions, namely the exponential
coupling i.e $h(\chi) = h_0e^{-\alpha\chi}$ and the cubic coupling
i.e $h(\chi) = h_0\big(\frac{\chi}{M}\big)^3$, recall that
$h(\chi) \sim \chi^2$ does not fit well with the Planck 2018
constraints and that is why we do not consider the quadratic
coupling in the present section. The exponential coupling along
with $H = H_0e^{-\Omega t}$ leads to the following effective
potential,
\begin{eqnarray}
 V_{eff}(\chi) = \frac{3H_0^2}{\kappa^2}e^{-2\Omega \chi/\mu^2} - \frac{2\Omega H_0}{\kappa^2}e^{-\Omega\chi/\mu^2}
 - 8h_0 \alpha H_0^3\mu^2 e^{-(3\Omega/\mu^2 + \alpha)\chi} - 24h_0 H_0^4 e^{-(4\Omega/\mu^2 +
 \alpha)\chi}\, ,
 \label{reheating5}
\end{eqnarray}
where we used the form of the function $\tilde{V}(\chi)$ as
obtained in Eq. (\ref{potential1}). Consequently the stable point
($<\chi>^{(ec)}$, where the notation ``ec'' stands for
``exponential coupling'') of $V_{eff}$ can be determined by the
following algebraic equation,
\begin{eqnarray}
 \frac{2\Omega H_0}{\kappa^2} - \frac{6H_0^2}{\kappa^2}e^{-\frac{\Omega}{\mu^2}<\chi>^{(ec)}}
 + 24 h_0H_0^3\alpha\mu^2 e^{-\frac{2\Omega}{\mu^2}<\chi>^{(ec)}} + 96 h_0H_0^4 e^{-\frac{3\Omega}{\mu^2}<\chi>^{(ec)}} =
 0\, .
 \label{reheating6}
\end{eqnarray}
The presence of $h_0$ in the above expression entails that the
Gauss-Bonnet coupling indeed affects the stability of the
inflaton. In order to understand the effect of the GB coupling
more clearly, we write $<\chi>^{(ec)} = <\chi>_0 + <\delta
\chi>^{(ec)}$, where $<\chi>_0$ is the stable point of $\chi$ in
absence of GB term ($h_0 = 0$) i.e.,
\begin{eqnarray}
 \frac{2\Omega H_0}{\kappa^2} - \frac{6H_0^2}{\kappa^2}e^{-\frac{\Omega}{\mu^2}<\chi>_0}&=&0\nonumber\\
 \Rightarrow
 e^{-\frac{\Omega}{\mu^2}<\chi>_0}&=&\frac{\Omega}{3H_0}\, .
 \label{reheating7}
\end{eqnarray}
Thus $<\delta \chi>^{(ec)}$ is the deviation of stable point from
$\chi>_0$ solely due to the presence of the Gauss-Bonnet term.
Expanding Eq. (\ref{reheating6}) in terms of $<\chi>^{(ec)} =
<\chi>_0 + <\delta \chi>^{(ec)}$, we get the following expression
for $<\delta \chi>^{(ec)}$,
\begin{eqnarray}
 <\delta \chi>^{(ec)} = -\frac{\mu^2}{\Omega} \frac{\frac{4}{3}h_0\kappa^2\Omega^2\big(\frac{4}{3} + \alpha\mu^2/\Omega\big)}
 {1 - \frac{8}{3}h_0\kappa^2\Omega^2\big(2 +
 \alpha\mu^2/\Omega\big)}\, ,
 \label{reheating8}
\end{eqnarray}
where we kept terms up to first order in $<\delta \chi>^{(ec)}$
and we also assumed $\frac{\Omega}{\alpha\mu^2} > 1$, which is
also consistent with the Planck observations, as mentioned earlier
in Section[\ref{sub1}]. Clearly $<\delta \chi>^{(ec)}$ becomes
zero as $h_0 \rightarrow 0$, as expected. Eqns. (\ref{reheating7})
and (\ref{reheating8}) immediately lead to the stable point of
$V_{eff}$ in presence of Gauss-Bonnet coupling, which is,
\begin{eqnarray}
 <\chi>^{(ec)}&=&<\chi>_0 + <\delta \chi>^{(ec)}\nonumber\\
 &=&\frac{\mu^2}{\Omega} \bigg[\ln{\big(3H_0/\Omega\big)} - \frac{\frac{4}{3}h_0\kappa^2\Omega^2\big(\frac{4}{3} + \alpha\mu^2/\Omega\big)}
 {1 - \frac{8}{3}h_0\kappa^2\Omega^2\big(2 +
 \alpha\mu^2/\Omega\big)}\bigg]\, .
 \label{reheating9}
\end{eqnarray}
Using the above expression for $<\chi>^{(ec)}$, we determine the
mass squared of the inflaton ($m^2_{(ec)}$) for the case of
exponential coupling function, which is,
\begin{eqnarray}
 m^2_{(ec)} = \frac{2\Omega^4}{\mu^4\kappa^2} \bigg[1 + \frac{\frac{8}{3}h_0\kappa^2\Omega^2\big(\frac{4}{3} + \alpha\mu^2/\Omega\big)
 \big(1 - 4h_0\kappa^2\Omega^2\big(\frac{4}{3} + \alpha\mu^2/\Omega\big)}
 {1 - \frac{8}{3}h_0\kappa^2\Omega^2\big(2 +
 \alpha\mu^2/\Omega\big)}\bigg]\, .
 \label{reheating10}
\end{eqnarray}
Thus in the absence of the Gauss-Bonnet term (i.e for $h_0 = 0$),
$m^2_{(ec)}$ becomes $m^2_{(ec)} =
\frac{2\Omega^4}{\mu^4\kappa^2}$ which is also consistent with Eq.
(\ref{reheating7}). However the presence of exponential coupling
affects the inflaton mass by the factor proportional to $h_0$, as
is evident from Eq. (\ref{reheating10}), in particular the mass
increases due to the presence of the Gauss-Bonnet term, compared
to the case where $h_0 = 0$. Having the explicit expression of
$m^2_{(ec)}$ (see Eq. (\ref{reheating10})) at hand, now we can
determine the reheating time by using Eq. (\ref{reheating4}),
which is,
\begin{eqnarray}
 t_h^{(ec)} = \frac{1}{\Omega} \ln{\bigg[\frac{8\pi
 m_{(ec)}H_0}{g^2\lambda^2}\bigg]}\, .
 \label{reheating11}
\end{eqnarray}
Thus we can argue that the presence of the exponential GB coupling
function, enhances the mass of the inflaton which in turn makes
the reheating time larger compared to the situation where the
Gauss-Bonnet term is absent.

For the cubic coupling ($h(\chi) = h_0\big(\chi/M\big)^3$), the
effective potential of the inflaton is equal to,
\begin{eqnarray}
 V_{eff}(\chi) = \frac{3H_0^2}{\kappa^2}e^{-2\Omega\chi/\mu^2} - \frac{2\Omega H_0}{\kappa^2}e^{-\Omega\chi/\mu^2}
 + \frac{24 h_0H_0^3\mu^2}{M^3}\chi^{2} e^{-3\Omega\chi/\mu^2} - \frac{24 h_0H_0^4}{M^3}\chi^3 e^{-4\Omega\chi/\mu^2}
 \label{reheating12}
\end{eqnarray}
Following the same procedure as above, we determine the stable
point of the effective potential and the mass of the inflaton
field, in the case of cubic coupling, which are,
\begin{eqnarray}
 <\chi>^{(cc)} = \frac{\mu^2}{\Omega} \bigg[\ln{\big(3H_0/\Omega\big)} -
 \frac{\frac{8}{3}h_0\kappa^2\Omega^2\big(\mu^2/\Omega M\big)^3\big(x_0 + x_0^2 + \frac{2}{3}x_0^3\big)}
 {1 - 4h_0\kappa^2\Omega^2\big(\mu^2/\Omega M\big)^3\big(\frac{4}{3}x_0 + x_0^2 -
 \frac{4}{3}x_0^3\big)}\bigg]\, ,
 \label{reheating13}
\end{eqnarray}
and
\begin{eqnarray}
 m^2_{(cc)} = \frac{2\Omega^4}{3\mu^4\kappa^2} \bigg[1 +
 \frac{\frac{16}{3}h_0\kappa^2\Omega^2\big(\mu^2/\Omega M\big)^3\big(x_0 + x_0^2 + \frac{2}{3}x_0^3\big)
 \bigg(1 + 12h_0\kappa^2\Omega^2\big(\mu^2/\Omega M\big)^3\big(x_0^2 - \frac{8}{9}x_0^3\big)\bigg)}
 {1 - 4h_0\kappa^2\Omega^2\big(\mu^2/\Omega M\big)^3\big(\frac{4}{3}x_0 + x_0^2 -
 \frac{4}{3}x_0^3\big)}\bigg]\, ,
 \label{reheating14}
\end{eqnarray}
respectively, where $x_0 = \ln{\big(3H_0/\Omega\big)}$ and the
notation 'cc' stands for ``cubic coupling``. Thereby, the presence
of cubic GB coupling function makes the inflaton mass larger
relative to the situation where the GB term is absent. As a
consequence, the reheating time $t_h^{(cc)} = \frac{1}{\Omega}
\ln{\bigg[\frac{8\pi m_{(cc)}H_0}{g^2\lambda^2}\bigg]}$ also
increases due to the effect of the cubic coupling function,
similar to the case of the exponential coupling we discussed
earlier.\\

Before closing, let us comment on an interesting issue, related to
previous works in the field. In Ref. \cite{new}, the authors
calculated the observational indices of inflation for a
generalized Galileon theories, however these theories are
quantitatively different to a great extent from the theory we
developed in this paper. Particularly, the theory at hand with
action (\ref{FRGBg22}) can be treated at a quantitative level as
an generalized Einstein-Gauss-Bonnet theory of gravity, which is
entirely different from the Galileon models studied in Ref.
\cite{new}. At a quantitative level, the theories developed in
Ref. \cite{new}, allow the derivation of general forms of the
observational indices, however in our case, and in Einstein
Gauss-Bonnet models, it is hard to derive general relations for
the observational indices. This is because the latter depend
strongly on the choice of the Gauss-Bonnet scalar coupling
function $h(\phi)$. Thus the results are strongly model dependent,
as we evinced in the previous sections, for example in Section VI
A with $h(\chi) = e^{-\alpha \chi}$ or in Section VI B with
$h(\chi) = \big(\chi/M\big)^n$. As we have shown, the quadratic
coupling is not viable although the exponential one in section VI
A is viable. We have further extended our discussion to
investigate the possible effects of GB coupling function on
reheating mechanism, unlike to \cite{new}.

\section{Conclusions \label{SecVII}}

In this work we studied the inflationary phenomenology of a
recently developed ghost-free $f(\mathcal{G})$ model of gravity.
Particularly, the form of the model mimics the scalar
Einstein-Gauss-Bonnet theory, so we employed the formalism of
cosmological perturbations for the latter theory, in order to
calculate the slow-roll indices and the corresponding
observational indices for the theory at hand. The model has rich
phenomenology due to the presence of a freely chosen function
$h(\chi)$, in which case by choosing this function and the Hubble
rate, the observational indices can be calculated easily. We
examined three types of inflationary cosmic evolution and
functional forms of the function $h(\chi)$, and as we demonstrated
it is possible to have a viable inflationary era, compatible with
the latest observational data. Particularly we used de Sitter,
quasi-de Sitter and exponential cosmological evolutions, and also
exponential and power-law functional forms for the function
$h(\chi)$. The simple de Sitter evolution leads in some cases to
problematic phenomenology, however no realistic cosmology gives
the exact de Sitter evolution, so we investigated the quasi-de
Sitter case, in which case the viability of the theory with the
observational data comes more easily. The same applies for the
exponential cosmological evolution. For the exponential Hubble
rate case, we also tested the stability of the first order scalar
and tensor cosmological perturbations, and as we demonstrated
these are stable for the same range of values of the free
parameters, for which the phenomenological viability of the model
is ensured. Finally we explore the reheating mechanism and the
possible effects of Gauss-Bonnet term on it for the case of
exponential Hubble rate. As a result we found that the presence of
GB coupling, in particular the exponential and cubic coupling
function, enhance the mass of the inflaton which in turn makes the
reheating time larger compared to the situation where the
Gauss-Bonnet term is absent. In this work we mainly focused on
realizing inflationary evolutions, however it is also possible to
realize non-singular cosmological evolutions, such as cosmological
bounces, however we defer this task to future work.

\section*{Acknowledgments}

This work is supported by MINECO (Spain), FIS2016-76363-P, and by
project 2017 SGR247 (AGAUR, Catalonia) (S.D.O). This work is also
supported by MEXT KAKENHI Grant-in-Aid for Scientific Research on
Innovative Areas ``Cosmic Acceleration'' No. 15H05890 (S.N.) and
the JSPS Grant-in-Aid for Scientific Research (C) No. 18K03615
(S.N.).


\begin{thebibliography}{99}


\bibitem{Guth:1980zm}
A.~H.~Guth,
Phys.\ Rev.\ D {\bf 23} (1981) 347. doi:10.1103/PhysRevD.23.347

\bibitem{Linde:1981mu}
A.~D.~Linde,
Phys.\ Lett.\  {\bf 108B} (1982) 389 [Adv.\ Ser.\ Astrophys.\
Cosmol.\  {\bf 3} (1987) 149]. doi:10.1016/0370-2693(82)91219-9



\bibitem{Albrecht:1982wi}
A.~Albrecht and P.~J.~Steinhardt,
Phys.\ Rev.\ Lett.\  {\bf 48} (1982) 1220 [Adv.\ Ser.\ Astrophys.\
Cosmol.\  {\bf 3} (1987) 158]. doi:10.1103/PhysRevLett.48.1220







\bibitem{Linde:2007fr}
A.~D.~Linde,
Lect.\ Notes Phys.\  {\bf 738} (2008) 1
doi:10.1007/978-3-540-74353-8\_1 [arXiv:0705.0164 [hep-th]].

\bibitem{Gorbunov:2011zzc}
D.~S.~Gorbunov and V.~A.~Rubakov,
Hackensack, USA: World Scientific (2011) 489 p doi:10.1142/7874


\bibitem{Lyth:1998xn}
D.~H.~Lyth and A.~Riotto,
Phys.\ Rept.\  {\bf 314} (1999) 1
doi:10.1016/S0370-1573(98)00128-8 [hep-ph/9807278].


\bibitem{Nojiri:2017ncd}
S.~Nojiri, S.~D.~Odintsov and V.~K.~Oikonomou,
Phys.\ Rept.\  {\bf 692} (2017) 1
doi:10.1016/j.physrep.2017.06.001 [arXiv:1705.11098 [gr-qc]].


\bibitem{Nojiri:2010wj}
S.~Nojiri and S.~D.~Odintsov,
Phys.\ Rept.\  {\bf 505} (2011) 59
doi:10.1016/j.physrep.2011.04.001 [arXiv:1011.0544 [gr-qc]].


\bibitem{Nojiri:2006ri}
S.~Nojiri and S.~D.~Odintsov,
eConf C {\bf 0602061} (2006) 06
 [Int.\ J.\ Geom.\ Meth.\ Mod.\ Phys.\  {\bf 4} (2007) 115]
doi:10.1142/S0219887807001928 [hep-th/0601213].


\bibitem{Capozziello:2011et}
S.~Capozziello and M.~De Laurentis,
Phys.\ Rept.\  {\bf 509} (2011) 167
doi:10.1016/j.physrep.2011.09.003 [arXiv:1108.6266 [gr-qc]].


\bibitem{Capozziello:2010zz}
V.~Faraoni and S.~Capozziello,
Fundam.\ Theor.\ Phys.\  {\bf 170} (2010).
doi:10.1007/978-94-007-0165-6


\bibitem{delaCruzDombriz:2012xy}
A.~de la Cruz-Dombriz and D.~Saez-Gomez,
Entropy {\bf 14} (2012) 1717 doi:10.3390/e14091717
[arXiv:1207.2663 [gr-qc]].


\bibitem{Olmo:2011uz}
G.~J.~Olmo,
Int.\ J.\ Mod.\ Phys.\ D {\bf 20} (2011) 413
doi:10.1142/S0218271811018925 [arXiv:1101.3864 [gr-qc]].






\bibitem{Starobinsky:1982ee}
A.~A.~Starobinsky,
Phys.\ Lett.\  {\bf 91B} (1980) 99.
doi:10.1016/0370-2693(80)90670-X




\bibitem{Nojiri:2003ft}
S.~Nojiri and S.~D.~Odintsov,
Phys.\ Rev.\ D {\bf 68} (2003) 123512
doi:10.1103/PhysRevD.68.123512 [hep-th/0307288].





\bibitem{Nojiri:2018ouv}
S.~Nojiri, S.~D.~Odintsov and V.~K.~Oikonomou,
arXiv:1811.07790 [gr-qc].



\bibitem{Akrami:2018odb}
Y.~Akrami {\it et al.} [Planck Collaboration],
arXiv:1807.06211 [astro-ph.CO].

\bibitem{Array:2015xqh}
P.~A.~R.~Ade {\it et al.} [BICEP2 and Keck Array Collaborations],
Phys.\ Rev.\ Lett.\ {\bf 116} (2016) 031302
doi:10.1103/PhysRevLett.116.031302 [arXiv:1510.09217
[astro-ph.CO]].




\bibitem{Nojiri:2006je}
S.~Nojiri, S.~D.~Odintsov and M.~Sami,
Phys.\ Rev.\ D {\bf 74} (2006) 046004
doi:10.1103/PhysRevD.74.046004 [hep-th/0605039].





\bibitem{Cognola:2006sp}
G.~Cognola, E.~Elizalde, S.~Nojiri, S.~Odintsov and S.~Zerbini,
Phys.\ Rev.\ D {\bf 75} (2007) 086002
doi:10.1103/PhysRevD.75.086002 [hep-th/0611198].



\bibitem{Nojiri:2005vv}
S.~Nojiri, S.~D.~Odintsov and M.~Sasaki,
Phys.\ Rev.\ D {\bf 71} (2005) 123509
doi:10.1103/PhysRevD.71.123509 [hep-th/0504052].


\bibitem{Nojiri:2005jg}
S.~Nojiri and S.~D.~Odintsov,
Phys.\ Lett.\ B {\bf 631} (2005) 1
doi:10.1016/j.physletb.2005.10.010 [hep-th/0508049].


\bibitem{Nojiri:2007te}
S.~Nojiri, S.~D.~Odintsov and P.~V.~Tretyakov,
Phys.\ Lett.\ B {\bf 651} (2007) 224
doi:10.1016/j.physletb.2007.06.029 [arXiv:0704.2520 [hep-th]].


\bibitem{Bamba:2014zoa}
K.~Bamba, A.~N.~Makarenko, A.~N.~Myagky and S.~D.~Odintsov,
JCAP {\bf 1504} (2015) 001 doi:10.1088/1475-7516/2015/04/001
[arXiv:1411.3852 [hep-th]].


\bibitem{Yi:2018gse}
Z.~Yi, Y.~Gong and M.~Sabir,
arXiv:1804.09116 [gr-qc].


\bibitem{Guo:2009uk}
Z.~K.~Guo and D.~J.~Schwarz,
Phys.\ Rev.\ D {\bf 80} (2009) 063523
doi:10.1103/PhysRevD.80.063523 [arXiv:0907.0427 [hep-th]].


\bibitem{Guo:2010jr}
Z.~K.~Guo and D.~J.~Schwarz,
Phys.\ Rev.\ D {\bf 81} (2010) 123520
doi:10.1103/PhysRevD.81.123520 [arXiv:1001.1897 [hep-th]].


\bibitem{Jiang:2013gza}
P.~X.~Jiang, J.~W.~Hu and Z.~K.~Guo,
Phys.\ Rev.\ D {\bf 88} (2013) 123508
doi:10.1103/PhysRevD.88.123508 [arXiv:1310.5579 [hep-th]].


\bibitem{Koh:2014bka}
S.~Koh, B.~H.~Lee, W.~Lee and G.~Tumurtushaa,
Phys.\ Rev.\ D {\bf 90} (2014) no.6,  063527
doi:10.1103/PhysRevD.90.063527 [arXiv:1404.6096 [gr-qc]].


\bibitem{Koh:2016abf}
S.~Koh, B.~H.~Lee and G.~Tumurtushaa,
Phys.\ Rev.\ D {\bf 95} (2017) no.12,  123509
doi:10.1103/PhysRevD.95.123509 [arXiv:1610.04360 [gr-qc]].


\bibitem{Kanti:2015pda}
P.~Kanti, R.~Gannouji and N.~Dadhich,
Phys.\ Rev.\ D {\bf 92} (2015) no.4,  041302
doi:10.1103/PhysRevD.92.041302 [arXiv:1503.01579 [hep-th]].


\bibitem{vandeBruck:2017voa}
C.~van de Bruck, K.~Dimopoulos, C.~Longden and C.~Owen,
arXiv:1707.06839 [astro-ph.CO].


\bibitem{Kanti:1998jd}
P.~Kanti, J.~Rizos and K.~Tamvakis,
Phys.\ Rev.\ D {\bf 59} (1999) 083512
doi:10.1103/PhysRevD.59.083512 [gr-qc/9806085].



\bibitem{Nozari:2017rta}
K.~Nozari and N.~Rashidi,
Phys.\ Rev.\ D {\bf 95} (2017) no.12,  123518
[arXiv:1705.02617 [astro-ph.CO]].

\bibitem{Chakraborty:2018scm}
S.~Chakraborty, T.~Paul and S.~SenGupta,
arXiv:1804.03004 [gr-qc].


\bibitem{Odintsov:2018zhw}
S.~D.~Odintsov and V.~K.~Oikonomou,
Phys.\ Rev.\ D {\bf 98} (2018) no.4,  044039
[arXiv:1808.05045 [gr-qc]].



\bibitem{Hwang:2005hb}
J.~c.~Hwang and H.~Noh,
Phys.\ Rev.\ D {\bf 71} (2005) 063536
[gr-qc/0412126].


\bibitem{Oikonomou:2018npe}
V.~K.~Oikonomou,
Phys.\ Rev.\ D {\bf 97} (2018) no.6,  064001
doi:10.1103/PhysRevD.97.064001 [arXiv:1801.03426 [gr-qc]].


\bibitem{Bamba:2014wda}
K.~Bamba, S.~Nojiri, S.~D.~Odintsov and D.~S\' aez-G\'omez,
Phys.\ Rev.\ D {\bf 90} (2014) 124061
doi:10.1103/PhysRevD.90.124061 [arXiv:1410.3993 [hep-th]].

\bibitem{Frampton:2011rh}
P.~H.~Frampton, K.~J.~Ludwick, S.~Nojiri, S.~D.~Odintsov and
R.~J.~Scherrer,
Phys.\ Lett.\ B {\bf 708} (2012) 204
doi:10.1016/j.physletb.2012.01.048 [arXiv:1108.0067 [hep-th]].


\bibitem{Liu:2012iba}
Z.~G.~Liu and Y.~S.~Piao,
Phys.\ Lett.\ B {\bf 713} (2012) 53
doi:10.1016/j.physletb.2012.05.027 [arXiv:1203.4901 [gr-qc]].


\bibitem{Kanti:2015dra}
P.~Kanti, R.~Gannouji and N.~Dadhich,
Phys.\ Rev.\ D {\bf 92} (2015) no.8,  083524
doi:10.1103/PhysRevD.92.083524 [arXiv:1506.04667 [hep-th]].

\bibitem{Calcagni:2006ye}
G.~Calcagni, B.~de Carlos and A.~De Felice,
Nucl.\ Phys.\ B {\bf 752} (2006) 404
doi:10.1016/j.nuclphysb.2006.06.020 [hep-th/0604201].

\bibitem{Kawai:1997mf}
S.~Kawai, M.~a.~Sakagami and J.~Soda,
gr-qc/9901065.


\bibitem{Soda:1998tr}
J.~Soda, M.~a.~Sakagami and S.~Kawai,
gr-qc/9807056.


\bibitem{Kawai:1998ab}
S.~Kawai, M.~a.~Sakagami and J.~Soda,
Phys.\ Lett.\ B {\bf 437} (1998) 284
doi:10.1016/S0370-2693(98)00925-3 [gr-qc/9802033].


\bibitem{Sberna:2017nzp}
L.~Sberna,
arXiv:1708.01150 [gr-qc].


\bibitem{Sberna:2017xqv}
L.~Sberna and P.~Pani,
Phys.\ Rev.\ D {\bf 96} (2017) no.12,  124022
doi:10.1103/PhysRevD.96.124022 [arXiv:1708.06371 [gr-qc]].



\bibitem{new}
T.~Kobayashi, M.~Yamaguchic and J.~Yokoyama
Prog.\ Theor. \ Phys. 126 (2011), 511-529
doi: 10.1143/PTP.126.511 [arXiv:1105.5723 [gr-qc]].





\end{thebibliography}
\end{document}